\let\includefigures=\iftrue

\input harvmac

\input epsf
%\def\epsfbox#1{\bigskip\centerline{$\bf\bullet\ #1\ \bullet$}\bigskip}

%%Last edited by ES, 9/7%%

\noblackbox
\def\IZ{\relax\ifmmode\mathchoice
{\hbox{\cmss Z\kern-.4em Z}}{\hbox{\cmss Z\kern-.4em Z}}
{\lower.9pt\hbox{\cmsss Z\kern-.4em Z}}
{\lower1.2pt\hbox{\cmsss Z\kern-.4em Z}}\else{\cmss Z\kern-.4em
Z}\fi}
\def\IB{\relax{\rm I\kern-.18em B}}
\def\IC{{\relax\hbox{\kern.3em{\cmss I}$\kern-.4em{\rm C}$}}}
\def\ID{\relax{\rm I\kern-.18em D}}
\def\IE{\relax{\rm I\kern-.18em E}}
\def\IF{\relax{\rm I\kern-.18em F}}
\def\IG{\relax\hbox{$\inbar\kern-.3em{\rm G}$}}
\def\IGa{\relax\hbox{${\rm I}\kern-.18em\Gamma$}}
\def\IH{\relax{\rm I\kern-.18em H}}
\def\II{\relax{\rm I\kern-.18em I}}
\def\IK{\relax{\rm I\kern-.18em K}}
\def\IP{\relax{\rm I\kern-.18em P}}
%\def\IX{\relax{\rm X\kern-.01em X}}
%this doesn't work

\def\p{\partial}

\font\cmss=cmss10 \font\cmsss=cmss10 at 7pt
\def\IR{\relax{\rm I\kern-.18em R}}

\def\ie{{\it i.e.}}
\def\frac#1#2{{#1 \over #2}}
\def\s{\sigma}

\def\OL#1{ \kern1pt\overline{\kern-1pt#1
   \kern-1pt}\kern1pt }

%\DouglasYP
\lref\DouglasYP{
M.~R.~Douglas, D.~Kabat, P.~Pouliot and S.~H.~Shenker,
``D-branes and short distances in string theory,''
Nucl.\ Phys.\ B {\bf 485}, 85 (1997)
[hep-th/9608024].
%%CITATION = HEP-TH 9608024;%%
}

%\MartinecCF
\lref\MartinecCF{
E.~J.~Martinec and W.~McElgin,
``String theory on AdS orbifolds,''
hep-th/0106171.
%%CITATION = HEP-TH 0106171;%%
}

%\BerkovitsHF
\lref\BerkovitsHF{
N.~Berkovits, A.~Sen and B.~Zwiebach,
``Tachyon condensation in superstring field theory,''
Nucl.\ Phys.\ B {\bf 587}, 147 (2000)
[hep-th/0002211].
%%CITATION = HEP-TH 0002211;%%
}

\lref\Marty{  
K. Bardakci, ``Dual Models and Spontaneous Symmetry Breaking", Nucl.Phys.
B68(1974)331;
K.Bardakci and M.B.Halpern, ``Explicit Spontaneous Breakdown in a Dual
Model", Phys.Rev.D10 (1974)4230;
K.Bardakci and M.B.Halpern, ``Explicit Spontaneous Breakdown in a Dual
Model II: N Point Functions", Nucl. Phys. B96(1975)285;
K.Bardakci, ``Spontaneous Symmetry Breakdown in the Standard Dual String
Model", Nucl.Phys.B133(1978)297.}

%\KosteleckyNT
\lref\KosteleckyNT{V.~A.~Kostelecky and S.~Samuel,
``On A Nonperturbative Vacuum For The Open Bosonic String,''
Nucl.\ Phys.\ B {\bf 336}, 263 (1990);
%%CITATION = NUPHA,B336,263;%%
}

\lref\BanksCH{
%\BanksCH
T.~Banks and L.~Susskind,
``Brane - Antibrane Forces,''
hep-th/9511194.
%%CITATION = HEP-TH 9511194;%%
}

%\SenMG
\lref\SenMG{A.~Sen,
``Non-BPS states and branes in string theory,''
hep-th/9904207.
%%CITATION = HEP-TH 9904207;%%
}

%\HarveyNA
\lref\HarveyNA{
J.~A.~Harvey, D.~Kutasov and E.~J.~Martinec,
cp ``On the relevance of tachyons,''
hep-th/0003101.
%%CITATION = HEP-TH 0003101;%%
}

%\GerasimovZP
\lref\GerasimovZP{A.~A.~Gerasimov and S.~L.~Shatashvili,
``On exact tachyon potential in open string field theory,''
JHEP {\bf 0010}, 034 (2000)
[hep-th/0009103].
%%CITATION = HEP-TH 0009103;%%
}

%\KutasovQP
\lref\KutasovQP{D.~Kutasov, M.~Marino and G.~Moore,
``Some exact results on tachyon condensation in string field theory,''
JHEP {\bf 0010}, 045 (2000)
[hep-th/0009148].
%%CITATION = HEP-TH 0009148;%%
}

%\GrossRK
\lref\GrossRK{D.~J.~Gross and W.~Taylor,
``Split string field theory. I,''
JHEP {\bf 0108}, 009 (2001)
[hep-th/0105059].
%%CITATION = HEP-TH 0105059;%%
}

%\RastelliUV
\lref\RastelliUV{L.~Rastelli, A.~Sen and B.~Zwiebach,
``Vacuum string field theory,''
hep-th/0106010.
%%CITATION = HEP-TH 0106010;%%
}

\lref\Opentach{
Owing to a lack of written reviews, for  reviews 
and further references we suggest\hfill\break
http://online.itp.ucsb.edu/online/mp01/sen1/ ,\hfill\break
http://www.slac.stanford.edu/spires/find/hep/www?rawcmd=ti+tachyon\%23 .
}

%\DouglasSW
\lref\DouglasSW{
M.~R.~Douglas and G.~Moore,
``D-branes, Quivers, and ALE Instantons,''
hep-th/9603167.
%%CITATION = HEP-TH 9603167;%%
}

%\OoguriWJ
\lref\OoguriWJ{
H.~Ooguri and C.~Vafa,
``Two-Dimensional Black Hole and Singularities of CY Manifolds,''
Nucl.\ Phys.\ B {\bf 463}, 55 (1996)
[hep-th/9511164].
%%CITATION = HEP-TH 9511164;%%
}

%\KachruYS
\lref\KachruYS{
S.~Kachru and E.~Silverstein,
``4d conformal theories and strings on orbifolds,''
Phys.\ Rev.\ Lett.\  {\bf 80}, 4855 (1998)
[hep-th/9802183].
%%CITATION = HEP-TH 9802183;%%
}

%\WittenGJ
\lref\WittenGJ{
E.~Witten,
``Instability Of The Kaluza-Klein Vacuum,''
Nucl.\ Phys.\ B {\bf 195}, 481 (1982).
%%CITATION = NUPHA,B195,481;%%
}

%\WittenZH
\lref\WittenZH{
E.~Witten,
``Some comments on string dynamics,''
hep-th/9507121.
%%CITATION = HEP-TH 9507121;%%
}

\lref\lebrun{
C. LeBrun, ``Counterexamples to the Generalized Positive
Action Conjecture'', Comm. Math. Phys. {\bf 118}, 591 (1988).
}

%\PolchinskiDY
\lref\PolchinskiDY{
J.~Polchinski,
``Scale And Conformal Invariance In Quantum Field Theory,''
Nucl.\ Phys.\ B {\bf 303}, 226 (1988).
%%CITATION = NUPHA,B303,226;%%
}

%\BanksYZ
\lref\BanksYZ{
T.~Banks and L.~J.~Dixon,
``Constraints On String Vacua With Space-Time Supersymmetry,''
Nucl.\ Phys.\ B {\bf 307}, 93 (1988).
%%CITATION = NUPHA,B307,93;%%
}

%\VafaIH
\lref\VafaIH{
C.~Vafa,
``Quantum Symmetries Of String Vacua,''
Mod.\ Phys.\ Lett.\ A {\bf 4}, 1615 (1989).
%%CITATION = MPLAE,A4,1615;%%
}

%\LiRP
\lref\LiRP{
M.~Li and T.~Yoneya,
``D-particle dynamics and the space-time uncertainty relation,''
Phys.\ Rev.\ Lett.\  {\bf 78}, 1219 (1997)
[hep-th/9611072].
%%CITATION = HEP-TH 9611072;%%
}

%\DeserTN
\lref\DeserTN{
S.~Deser, R.~Jackiw and G.~'t Hooft,
``Three-Dimensional Einstein Gravity: Dynamics Of Flat Space,''
Annals Phys.\  {\bf 152}, 220 (1984).
%%CITATION = APNYA,152,220;%%
}

%\CostaNW
\lref\CostaNW{
M.~S.~Costa and M.~Gutperle,
``The Kaluza-Klein Melvin solution in M-theory,''
JHEP {\bf 0103}, 027 (2001)
[hep-th/0012072].
%%CITATION = HEP-TH 0012072;%%
}

%\GutperleMB
\lref\GutperleMB{
M.~Gutperle and A.~Strominger,
``Fluxbranes in string theory,''
JHEP {\bf 0106}, 035 (2001)
[hep-th/0104136].
%%CITATION = HEP-TH 0104136;%%
}

%\SilversteinNS
\lref\SilversteinNS{
E.~Silverstein and Y.~S.~Song,
``On the critical behavior of D1-brane theories,''
JHEP {\bf 0003}, 029 (2000)
[hep-th/9912244].
%%CITATION = HEP-TH 9912244;%%
}

%\AdamsJB
\lref\AdamsJB{
A.~Adams and E.~Silverstein,
``Closed string tachyons, AdS/CFT, and large N QCD,''
hep-th/0103220.
%%CITATION = HEP-TH 0103220;%%
}

%\WittenYC
\lref\WittenYC{
E.~Witten,
``Phases of N = 2 theories in two dimensions,''
Nucl.\ Phys.\ B {\bf 403}, 159 (1993)
[hep-th/9301042].
%%CITATION = HEP-TH 9301042;%%
}

%\Buscher
\lref\Buscher{
T.~H.~Buscher,
``A Symmetry Of The String Background Field Equations,''
Phys.\ Lett.\ B {\bf 194}, 59 (1987);
%%CITATION = PHLTA,B194,59;%%
``Path Integral Derivation Of Quantum Duality In Nonlinear Sigma Models,''
Phys.\ Lett.\ B {\bf 201}, 466 (1988).
%%CITATION = PHLTA,B201,466;%%
}

%\TseytlinII
\lref\TseytlinII{
A.~A.~Tseytlin and K.~Zarembo,
``Effective potential in non-supersymmetric SU(N) x SU(N) gauge theory  and
interactions of type 0 D3-branes,''
Phys.\ Lett.\ B {\bf 457}, 77 (1999)
[hep-th/9902095].
%%CITATION = HEP-TH 9902095;%%
}

%\DowkerUP
\lref\DowkerUP{
F.~Dowker, J.~P.~Gauntlett, S.~B.~Giddings and G.~T.~Horowitz,
%``On pair creation of extremal black holes and Kaluza-Klein monopoles,''
Phys.\ Rev.\ D {\bf 50}, 2662 (1994)
[hep-th/9312172].
%%CITATION = HEP-TH 9312172;%%
}

%\DowkerBT
\lref\DowkerBT{
F.~Dowker, J.~P.~Gauntlett, D.~A.~Kastor and J.~Traschen,
%``Pair creation of dilaton black holes,''
Phys.\ Rev.\ D {\bf 49}, 2909 (1994)
[hep-th/9309075].
%%CITATION = HEP-TH 9309075;%%
}

%\DowkerGB
\lref\DowkerGB{
F.~Dowker, J.~P.~Gauntlett, G.~W.~Gibbons and G.~T.~Horowitz,
``The Decay of magnetic fields in Kaluza-Klein theory,''
Phys.\ Rev.\ D {\bf 52}, 6929 (1995)
[hep-th/9507143].
%%CITATION = HEP-TH 9507143;%%
}

%\ZamolodchikovGT
\lref\ZamolodchikovGT{
A.~B.~Zamolodchikov,
``Irreversibility Of The Flux Of The Renormalization Group In A 2-D Field
Theory,'' JETP Lett.\  {\bf 43}, 730 (1986)
[Pisma Zh.\ Eksp.\ Teor.\ Fiz.\  {\bf 43}, 565 (1986)].
%%CITATION = JTPLA,43,730;%%
}

%\deAlwisPR
\lref\deAlwisPR{
S.~P.~de Alwis, J.~Polchinski and R.~Schimmrigk,
``Heterotic Strings With Tree Level Cosmological Constant,''
Phys.\ Lett.\ B {\bf 218}, 449 (1989).
%%CITATION = PHLTA,B218,449;%%
}

%\GregoryTE
\lref\GregoryTE{
R.~Gregory, J.~A.~Harvey and G.~Moore,
``Unwinding strings and T-duality of Kaluza-Klein and H-monopoles,''
Adv.\ Theor.\ Math.\ Phys.\  {\bf 1}, 283 (1997)
[hep-th/9708086].
%%CITATION = HEP-TH 9708086;%%
}

%\KachruED
\lref\KachruED{
S.~Kachru, J.~Kumar and E.~Silverstein,
``Orientifolds, RG flows, and closed string tachyons,''
Class.\ Quant.\ Grav.\  {\bf 17}, 1139 (2000)
[hep-th/9907038].
%%CITATION = HEP-TH 9907038;%%
}

%\HorowitzGN
\lref\HorowitzGN{
G.~T.~Horowitz and L.~Susskind,
``Bosonic M theory,''
hep-th/0012037.
%%CITATION = HEP-TH 0012037;%%
}

%\FabingerJD
\lref\FabingerJD{
M.~Fabinger and P.~Horava,
``Casimir effect between world-branes in heterotic M-theory,''
Nucl.\ Phys.\ B {\bf 580}, 243 (2000)
[hep-th/0002073].
%%CITATION = HEP-TH 0002073;%%
}

%\WittenCD
\lref\WittenCD{
E.~Witten,
``D-branes and K-theory,''
JHEP {\bf 9812}, 019 (1998)
[hep-th/9810188].
%%CITATION = HEP-TH 9810188;%%
}

%\EllwoodPY
\lref\EllwoodPY{
I.~Ellwood and W.~Taylor,
``Open string field theory without open strings,''
Phys.\ Lett.\ B {\bf 512}, 181 (2001)
[hep-th/0103085].
%%CITATION = HEP-TH 0103085;%%
}

%\YiHD
\lref\YiHD{
P.~Yi,
``Membranes from five-branes and fundamental strings from Dp branes,''
Nucl.\ Phys.\ B {\bf 550}, 214 (1999)
[hep-th/9901159].
%%CITATION = HEP-TH 9901159;%%
}

%\BergmanKM
\lref\BergmanKM{
O.~Bergman and M.~R.~Gaberdiel,
``Dualities of type 0 strings,''
JHEP {\bf 9907}, 022 (1999)
[hep-th/9906055].
%%CITATION = HEP-TH 9906055;%%
}

%\JohnsonCH
\lref\JohnsonCH{
C.~V.~Johnson,
``D-brane primer,''
hep-th/0007170.
%%CITATION = HEP-TH 0007170;%%
}

%\AnselmiSM
\lref\AnselmiSM{
D.~Anselmi, M.~Billo, P.~Fre, L.~Girardello and A.~Zaffaroni,
``ALE manifolds and conformal field theories,''
Int.\ J.\ Mod.\ Phys.\ A {\bf 9}, 3007 (1994)
[hep-th/9304135].
%%CITATION = HEP-TH 9304135;%%
}

%\SrednickiMQ
\lref\SrednickiMQ{
M.~Srednicki,
``IIB or not IIB,''
JHEP {\bf 9808}, 005 (1998)
[hep-th/9807138].
%%CITATION = HEP-TH 9807138;%%
}

\lref\horowitz{G. Horowitz, work in progress}

%\WittenMF
\lref\WittenMF{
E.~Witten,
``A Simple Proof Of The Positive Energy Theorem,''
Commun.\ Math.\ Phys.\  {\bf 80}, 381 (1981).
%%CITATION = CMPHA,80,381;%%
}

%\GibbonsGU
\lref\GibbonsGU{
G.~W.~Gibbons and C.~N.~Pope,
``Positive Action Theorems For Ale And Alf Spaces,''
ICTP-81-82-20.
}

%\EllwoodIG
\lref\EllwoodIG{
I.~Ellwood, B.~Feng, Y.~He and N.~Moeller,
``The identity string field and the tachyon vacuum,''
JHEP {\bf 0107}, 016 (2001)
[hep-th/0105024].
%%CITATION = HEP-TH 0105024;%%
}

%\CostaIF
\lref\CostaIF{
M.~S.~Costa, C.~A.~Herdeiro and L.~Cornalba,
``Flux-branes and the dielectric effect in string theory,''
hep-th/0105023.
%%CITATION = HEP-TH 0105023;%%
}

\lref\SaffinKY{
P.~M.~Saffin,
``Gravitating fluxbranes,''
Phys.\ Rev.\ D {\bf 64}, 024014 (2001)
[gr-qc/0104014].
%%CITATION = GR-QC 0104014;%%
}

\lref\uranga{
A. Uranga, work in progress; K. Dasgupta, private communication.
}

%\DabholkarAI
\lref\DabholkarAI{
A.~Dabholkar,
``Strings on a cone and black hole entropy,''
Nucl.\ Phys.\ B {\bf 439}, 650 (1995)
[hep-th/9408098].
%%CITATION = HEP-TH 9408098;%%
}

%\DabholkarGG
\lref\DabholkarGG{
A.~Dabholkar,
``Quantum corrections to black hole entropy in string theory,''
Phys.\ Lett.\ B {\bf 347}, 222 (1995)
[hep-th/9409158].
%%CITATION = HEP-TH 9409158;%%
}

%\LoweAH
\lref\LoweAH{
D.~A.~Lowe and A.~Strominger,
``Strings near a Rindler or black hole horizon,''
Phys.\ Rev.\ D {\bf 51}, 1793 (1995)
[hep-th/9410215].
%%CITATION = HEP-TH 9410215;%%
}

%\DixonJW
\lref\DixonJW{
L.~J.~Dixon, J.~A.~Harvey, C.~Vafa and E.~Witten,
``Strings On Orbifolds,''
Nucl.\ Phys.\ B {\bf 261}, 678 (1985).
%%CITATION = NUPHA,B261,678;%%
}

%\DixonJC
\lref\DixonJC{
L.~J.~Dixon, J.~A.~Harvey, C.~Vafa and E.~Witten,
``Strings On Orbifolds. 2,''
Nucl.\ Phys.\ B {\bf 274}, 285 (1986).
%%CITATION = NUPHA,B274,285;%%
}

\Title{\vbox{\baselineskip12pt\hbox{hep-th/0108075}
\hbox{NSF-ITP-01-75}
\hbox{SLAC-PUB-8955}
}}
{\vbox{
{\centerline{Don't Panic!
Closed String Tachyons in ALE Spacetimes}}
}
}

%\Title{\vbox{\baselineskip12pt}}
%{\vbox{
%        \centerline{Don't Panic! Closed String Tachyons in ALE
%Spacetimes}
%       \centerline{Closed String Tachyon Condensation to ALE Spacetimes}
%       \centerline{Quivering Tachyonic Orbifolds}
%       \centerline{Auto-Enzymatic Quiver Introns and D-RNA}
%       \centerline{NS5-Branes, Quivering ALE
%Instantons}\medskip\centerline{and Closed String Tachyons }
%       \centerline{A La Recherche du SUSY Perdu}
%       \centerline{A La Recherche du Tachyons Perdu}
%       \centerline{Breaking SUSY is Hard to Do}
%       \centerline{There and Back Again: Closed String Tachyon
%Condensation in Type II Orbifolds}
%       \centerline{On the Persistence of SUSY}
%       \centerline{The Rise and Fall of the Closed String Tachyon}
%       \centerline{Three Brains Probing Tachyonic Orbifolds}
%       \centerline{Closed String Tachyons, AFL/CIO, and Large-$\eta$ CCDs}
%       \centerline{Faster, Tachyon, Kill, Kill!}
%       \centerline{The Once and Future SUSY}

\medskip
\centerline{A. Adams$^{1,2}$, J. Polchinski$^{2}$ and E. Silverstein$^{1,2}$}
\medskip
\centerline{$^{1}$Department of Physics and SLAC}
\centerline{Stanford University}
\centerline{Stanford, CA 94305/94309}
\medskip
\centerline{$^{2}$Institute for Theoretical Physics}
\centerline{University of California}
\centerline{Santa Barbara, CA 93106}
%\centerline{$^{3}$A Hospital in Tunisia, I Fear...}
\bigskip
\noindent

% ABSTRACT

We consider closed string tachyons localized at the fixed points of
noncompact nonsupersymmetric orbifolds.
We argue that tachyon condensation drives these
orbifolds to flat space or supersymmetric ALE
spaces.  The decay proceeds via an expanding shell of dilaton gradients
and curvature which interpolates between two regions of distinct angular
geometry.  The string coupling remains weak throughout.  For small tachyon
VEVs, evidence comes from quiver theories on D-branes probes, in which
deformations by twisted couplings smoothly connect non-supersymmetric
orbifolds to supersymmetric orbifolds of reduced order.  For large tachyon
VEVs, evidence comes from worldsheet RG flow and spacetime gravity.  For
$\IC^2/\IZ_n$, we exhibit infinite sequences of transitions producing SUSY ALE
spaces via twisted closed string  condensation from non-supersymmetric ALE
spaces. In a $T$-dual description this provides a mechanism for creating
NS5-branes via {\it closed} string tachyon condensation similar to the
creation of D-branes via {\it open} string tachyon condensation.  We also
apply our results to recent duality conjectures involving fluxbranes and the
type 0 string.

\Date{August 2001}
%\draftmode

\baselineskip=18pt

\newsec{Motivation and Outline}

An understanding of the vacuum structure of String/M theory after
supersymmetry breaking is crucial for phenomenology and
cosmology.  It is also relevant to the question of unification;
it is important to understand the extent and nature of
connections between different vacua in the theory.
A basic issue is the fate of theories that have tachyons in their
tree-level spectra.  This has long been a source of puzzlement, but for open
strings there has been a great deal of progress.\foot{
A small sampling of references on this subject is  
\Marty\KosteleckyNT\BanksCH\SenMG\HarveyNA\GerasimovZP\KutasovQP
\GrossRK\RastelliUV\Opentach.}  
Open string tachyons generally have an interpretation in terms
of D-brane annihilation, binding, or decay, and a quantitative description
of these processes has been achieved by an assortment of methods from
conformal field theory, string field theory, and noncommutative geometry.
This has also led to a deeper understanding of the role of K
theory, and the reanimation of open string field theory.

For closed string tachyons the understanding is much more rudimentary.
These should be connected with the decay of spacetime itself,
rather than of branes in embedded in a fixed spacetime.  In this paper we
study a class of tachyonic closed string theories in which the decay can be
followed with reasonable confidence.  The key feature of these theories is
that the bulk of spacetime is stable, and the tachyons live only on a
submanifold.  Thus they are similar to the tachyonic open string theories,
and we will note certain close parallels,
though in the absence of closed string field theory
we will not be able to achieve as
complete a quantitative control.

The theories we study are noncompact, nonsupersymmetric orbifolds 
\DixonJW\DixonJC\ of
ten-dimensional superstring theories 
\DabholkarAI\DabholkarGG\LoweAH\AdamsJB\MartinecCF.
The simplest case is to identify two
dimensions under a rotation by $2\pi/n$, forming a cone with deficit angle
$2\pi-2\pi/n$
($n$ must be odd, for reasons to be explained in \S2).
This is the simplest example of an Asymptotically Locally Euclidean (ALE)
space, which is defined generally as any space whose geometry at long
distance is of the form $\IR^k/\Gamma$, with $\Gamma$ some subgroup of the
rotation group.
The tip of the cone, which is singular, is a seven-dimensional submanifold.
The rotation leaves no spinor invariant and so supersymmetry is completely
broken, and there are tachyons in the twisted sector of the orbifold theory.
Where do these tachyons take us?

There are several plausible guesses, based on experience in other
systems:\hfill\break (I) A hole might appear at the tip, and then expand to
consume spacetime. Such a reduction in degrees of freedom
is naively suggested by the relevance of the tachyon
vertex operator at zero momentum \deAlwisPR, and by the presence of a
nonperturbative Kaluza-Klein instability in certain backgrounds \WittenGJ,
and has been argued to be the fate of other tachyonic closed string
theories \KachruED\FabingerJD
\HorowitzGN\CostaNW\GutperleMB.\hfill\break  (II) The tip might
begin to elongate, asymptotically approaching the infinite throat geometry
that is often found in singular conformal field theories
\WittenZH.\hfill\break  (III) The tip might smooth out, by analogy to
the effect of the marginal twisted sector perturbations in supersymmetric
orbifolds.  This smoothing might stop at the string scale, or continue
indefinitely.

We will argue that it is the last of these that occurs, as
was also suggested recently in \MartinecCF.  At late times,
when a general relativistic analysis is valid, an expanding dilaton pulse
travels outward with the speed of light.  This is depicted in figure~1.
The
region interior to the pulse is flat, with vanishing deficit angle.
The energy
contained within the pulse produces the jump to
the asymptotic deficit angle of $2\pi-2\pi/n$ \DeserTN.
More generally, by following special directions in the space of tachyons,
the decay can take place in a series of steps, where for example
$\IC/\IZ_{2l+1}$ decays via a dilaton pulse to $\IC/\IZ_{2l-1}$, or to
any $\IC/\IZ_{2l^\prime+1}$ orbifold with $l^\prime <l$.\bigskip

\epsfbox{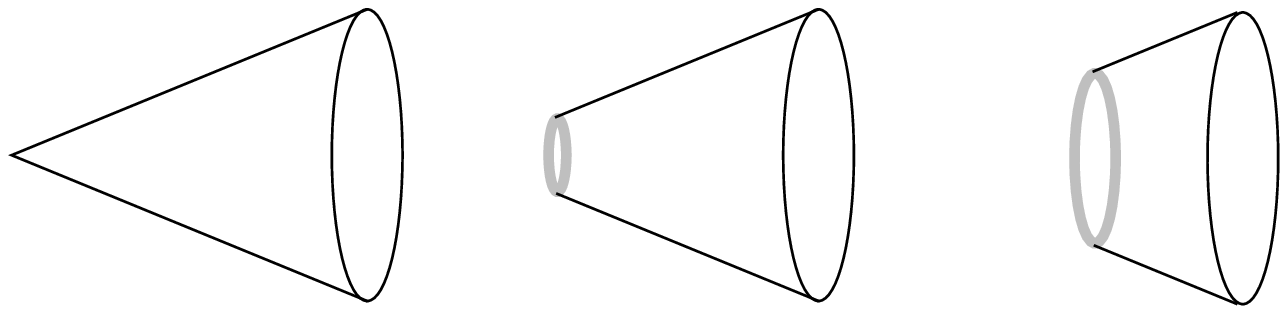}
\smallskip\noindent
Figure 1: Decay of the conic singularity.  The end of the cone
is replaced by a flat base.  The outward-moving dilaton pulse is
shown in gray.
\bigskip

We will analyze this process in two complementary regimes.  When the
tachyon expectation value is small and so the smoothed region small compared
to the string scale, we use D-brane probes~\DouglasYP, whose world-volume
theory is a quiver gauge theory \DouglasSW.  This is the {\it substring
regime}. D-branes on supersymmetric orbifolds have
been studied extensively; we extend these techniques to study
non-supersymmetric orbifold compactifications with closed string tachyons.
The probes  see a smoothed geometry when the tachyon is nonzero.  When the
smoothing region exceeds the string length scale, we can instead use a general
relativistic description, and we argue that the solution has the form in
figure~1.  This is the {\it gravity regime}.  Because
of $\alpha^\prime$ corrections, we do not have a controlled
approximation that connects the two regimes, but we argue that together they
give a simple and consistent picture of a transition from a conic singularity
to flat space via tachyon condensation.  These same complementary descriptions
have been applied to type I instantons and to supersymmetric ALE spaces
\DouglasSW.

If one or more of these singularities is part of a compact manifold, then
the initial stages of tachyon condensation will be the same near each orbifold
fixed point, producing a smooth compact geometry.  Unlike the noncompact
system, which evolves forever, we will argue that the compact space collapses
toward a Big Crunch in finite time.

If we put $N$ D3-branes on the fixed plane, and consider
the near horizon limit, then the system
is expected to be dual to a nonsupersymmetric gauge theory \KachruYS.  The
fixed plane is partly transverse to the D3-brane, and in the large-radius
limit of the $AdS_5\times S^5/\IZ_n$ background we find a dramatic instability
that grows toward the boundary of AdS.  We argue that at large 't Hooft
coupling these nonsupersymmetric field theories are unphysical.
This contrasts with the more benign
infrared Coleman-Weinberg instabilities evident on the field theory side at
weak 't Hooft coupling (small radius), which arise in theories whose gravity
duals have tachyons at large radius
\AdamsJB\TseytlinII.

The analysis can also be applied to orbifolds the form $\IC^q/\Gamma$, where
$\Gamma$ is a discrete subgroup of the rotation group that fixes a point in
$\IC^q$.  If
$\Gamma\not\subset SU(q)$ then the background does not preserve
supersymmetry, and there are twisted-sector tachyons localized at the fixed
point.  For $q=1$ there is no $\Gamma$ that preserves supersymmetry, but for
$q \geq 2$ there will be.

We study several $\IC^2/\IZ_n$ examples, where analysis
of the quiver theories on D-brane probes leads to predictions for
transitions between different orbifolds.  There is a new
effect that can occur in this case: we exhibit infinite sequences of examples
with transitions from {\it nonsupersymmetric} ALE spaces to {\it
supersymmetric} ones.  We again expect a gravitational background for large
tachyon VEV that involves an expanding shell of dilaton gradient, combined
with metric curvature.  In this case the total energy of the transition region
must vanish, since both the initial and final ALE spaces have vanishing energy
as measured by the falloff of the gravitational fields at infinity.  (We will
explain why this is not inconsistent with positive energy theorems.)

For large $n$ in $\Gamma=\IZ_n$, an angular direction is small for a
significant range of radii and it is useful to go to a $T$-dual picture
involving NS5-branes \OoguriWJ. These
transitions therefore provide a closed string analogue of the open
string brane descent relations \BerkovitsHF, in that we can realize for
example any $A_k$ space (and therefore the equivalent dual system of $k+1$
NS5-branes) via tachyon condensation (and/or marginal deformation)
from a non-supersymmetric ALE space.
Also, by adding an R-R flux, which
has little effect on the geometry, we obtain a system which has a
conjectured dual
description in terms of fluxbranes~\DowkerBT\DowkerUP\CostaNW\GutperleMB
\CostaIF\SaffinKY.

This realization of supersymmetric ALE spaces (and therefore NS branes)
by closed-string tachyon condensation is very reminiscent of
similar constructions in open string theory \Opentach.  In this
regard, we should emphasize that certain puzzles that arise in the open string
case arise here as well \MartinecCF.\foot{We thank
M. Berkooz, P. Kraus, E. Martinec, and other participants of
the Amsterdam Summer workshop for discussions on this.}
In particular, in open string tachyon
condensation, one finds gauge fields without sufficient
perturbative charged
matter to Higgs them \SrednickiMQ\WittenCD, but open string field
theory calculations at disk order suggest that they are nonetheless lifted
classically \Opentach\EllwoodPY\EllwoodIG.  In the tachyonic closed string
systems we study here, there are twisted RR gauge potentials
without perturbative charged matter; our evidence suggests
that the defect is nonetheless smoothed and the RR potential
lifted.  In both the open and closed string cases, it would
be very interesting to understand the classical stringy
effect, evidently going beyond ordinary effective quantum
field theory, which allows gauge fields to be so lifted.
In the closed string case, it would be interesting to
understand an analogue of the quantum confinement effect
identified in the open string case in \YiHD; here as
there one has D-branes charged under the gauge group
of interest (in our case these are the fractional D-branes,
whose condensation might lead to confinement of twisted
strings into untwisted strings).

The organization of this paper is as follows.  In \S2\ we discuss the
$\IC/\IZ_n$ orbifold, including the twisted
sector spectrum and the quantum symmetry group of
the orbifold theory.  We also discuss the difference between orbifolds and
ALE spaces that do not have orbifold descriptions.  We consider D-brane probes
in the orbifold theory, deriving the quiver representation.
In \S3 we analyze the
quiver theory/linear sigma model for the $\IC/\IZ_n$ orbifolds and their
twisted deformations, discussing both generic decays and decays
that leave lower-order orbifolds.
In \S4 we
develop the general relativistic description of these same solutions.  We
discuss renormalization group evolution and time evolution.  These are
similar, in that both lead to a smoothed region that grows without bound, but
there are differences in the details.  We discuss the consistency between the
renormalization group analysis and the
$c$-theorem.  We then discuss the fate of compact spaces with
nonsupersymmetric orbifold points, and the consequences of our results for
AdS/CFT duality.
In \S5, we analyze transitions in the $\IC^2/\Gamma$ case by means of the
quiver theories, and exhibit decays from non-SUSY to SUSY ALE spaces.  In
the general relativistic regime we explain how our results are consistent with
positive energy theorems.  Finally, in \S6\ we discuss dual systems,
including fluxbranes, and in \S7\ mention some directions for further research.

\newsec{The $\IC/\IZ_n$ Orbifold}

\subsec{Closed String Spectrum}

Let us start by reviewing some of the basic properties of
the $\IC/\IZ_n$ orbifold conformal field theory 
\DabholkarAI\DabholkarGG\LoweAH.  
These orbifolds are defined by identifying the 8-9 plane under a rotation $R$
through $2\pi/n$.  This allows two possible actions on the spinors,
\eqn\rotations{
R = \exp( {2\pi i} J_{89} /n )\quad \hbox{or}\quad
\exp( {2\pi i} J_{89} )\exp( {2\pi i} J_{89} /n )\ ,
}
where $J_{89}$ is the rotation generator.  For either choice, $R^n$ acts
trivially on spacetime and so is either 1 or $\exp( {2\pi i} J_{89} )
= (-1)^{\bf F}$.  If
$R^n = (-1)^{\bf F}$,
the orbifold group (which is actually $\IZ_{2n}$ in this case) includes
this operator and so projects out spacetime fermions and introduces
tachyons in the bulk.  Because we want to have all tachyons localized at the
fixed point we must have
$R^n = 1$.  For the two choices \rotations\ one finds
\eqn\rton{
R^n = (-1)^{\bf F}\quad \hbox{or}\quad
(-1)^{(n+1)\bf F}\ .
}
Thus, only the second choice of $R$ is acceptable, and only for $n$ odd:
\eqn\goodrot{
R =
\exp\biggl( {2\pi i}\frac{n+1}{n} J_{89} \biggr)\ ,\quad n = 2l+1\ .
 }

In the sector twisted by $R^k$ ($1 \leq k \leq n-1$), in the
light-cone Green-Schwarz description there are six real untwisted scalars,
one complex
scalar twisted by $k/n$, and four complex
fermions twisted by $k/2 + k/2n$.  The
standard calculation of the zero-point energy gives
\eqn\zpe{
\frac{\alpha'}{4} m^2 = \left\{ {-k/2n\ ,\ k\ \hbox{even}\ ,}\atop{ (k-n)/2n
\ ,\ k\ \hbox{odd} \ .}\right.
}
Thus the lowest state is tachyonic in every twisted sector.  There are also
excited state tachyons in many sectors.  For example, when $k=1$ the lowest
twisted scalar excitation takes $(1-n)/2n$ to $(3-n)/2n$ and so this state is
tachyonic for $n > 3$.  Our analysis will be rather coarse, and so we will
generally not distinguish the ground state in each sector from excited states
with the same symmetries.

We wish to ask, where do these tachyonic perturbations take the system?
Since they are in the twisted sectors, their initial effect is in the
neighborhood of the fixed point.  There are two contexts to consider.  First,
we could add the tachyonic vertex operators to the Hamiltonian.  Tachyonic
states correspond to relevant vertex operators, so they change the IR behavior
of the world-sheet theory.  We are then interested in determining the
renormalization group (RG) flow.  Second, we could consider a time-dependent
string solution that begins as a small but exponentially growing tachyonic
perturbation of the orbifold.  We are then interested in the subsequent
time evolution.

Physically these are distinct questions.  The first is an
off-shell question from the point of view of string theory, but well posed as
question in two-dimensional quantum field theory.  The second is an
on-shell question in string theory.  In fact we will find, as has been seen in
other contexts, that the scale and time evolutions are similar.  In both cases
the question can be posed in the classical string limit, with no string loop
effects.  If the world-sheet theory were to become singular, for example if
the dilaton were to become large, then this framework would break down,
but we will find that at least generically this does not happen.

The orbifold preserves an $SO(7,1) \times U(1)$ subgroup of the parent
$SO(9,1)$.  In addition there is a new ``quantum'' symmetry that appears in
the orbifolded theory \VafaIH: the twist is conserved, mod $n$.\foot{This
is not the same spacetime $\IZ_n$ group used to construct the orbifold.  All
states are invariant by definition under that symmetry, while the quantum
$\IZ_n$ is carried by the twisted sector states.} The lowest tachyons in
general break the quantum symmetry completely but leave the $SO(7,1) \times
U(1)$ unbroken (in the RG case) or break it to $SO(7) \times U(1)$ (in the
time-dependent case).  In some cases we will consider perturbations that
leave part of the quantum symmetry unbroken, while in others we will find that
a new quantum symmetry, unrelated to the original one, emerges asymptotically.

Actually, the evolution is more restricted than would follow from spacetime
symmetry alone.  The $X^M$ and $\psi^M$ (of the RNS description) upon which the
$SO(7,1)$ or $SO(7)$ acts do not appear in the perturbation.  These fields
therefore remain free, whereas the symmetry would allow a warp factor
depending on the other coordinates.

We will consider processes where a $\IZ_{2l+1}$ singularity emits a radiation
pulse with just the appropriate energy to leave behind a $\IZ_{2l-1}$
singularity.  We could also imagine the time-reversed process, sending in a
pulse with the appropriate energy.

This raises an interesting issue.  We have found that there are no
$\IZ_{2l}$ orbifolds with supersymmetry broken only locally
at the tip of the cone, but what if we consider a solution
of type II string theory which describes a pulse sent inward
with just the right
energy to create such a singularity?  When the pulse reaches the origin, the
geometry is a cone with deficit angle $2\pi - 2\pi/2l$.  The difference
between this case and the time reversal of our orbifold decay
process is
that here there is no simple description of the singularity.  Away from the
singularity, the lines $\theta = 0$ and $\theta = 2\pi / 2l$ are identified
under a rotation $\exp( {2\pi i} J_{89} /2l )$ or $\exp( {2\pi i} J_{89}
)\exp( {2\pi i} J_{89} /2l)$.  This is a sensible configuration, and the
dynamical process allows one to reach it.  However, on the $2l$-fold covering
space, the lines $\theta = 0$ and $\theta = 2\pi$ are identified under the
action of $(-1)^{\bf F}$, and so there is a branch cut in the spinor fields.
This is the essential difference from the orbifold: in the orbifold the
untwisted fields are single-valued on the covering space.  We could similarly
consider a wedge of any opening angle $\theta_{\rm o}$, where the plane is
generically not a covering space.  Again, dynamically we could construct a
state that has this behavior away from the singularity, but that
within a string
distance of the singularity has some complicated description, not based on a
free \hbox{CFT}, if the singularity is resolved at all.
Indeed, we will find many example of orbifolds decaying to
such spaces; we will use the terms `quasi-orbifold' or `quasi-ALE (QALE)' to
refer to these more general spaces that are locally Euclidean but are not
obtained as orbifolds of a single-valued theory on Euclidean space.

\subsec{Open String Spectrum}

We now consider a D$p$-brane probe of the geometry.  Here as in many
other contexts, D-brane probes \DouglasSW\ and closely related linear
sigma model techniques \WittenYC\ are useful for obtaining a broader
view of the space of closed string backgrounds than is available
from perturbation theory about a specific world-sheet
CFT.  In studying a D-brane probe, the low energy quantum field
theory on its world-volume is only valid in the
substring regime, where the VEVs of world-volume scalars
(scaled to have dimensions of length) are sufficiently small compared
to the string length $\sqrt{\alpha^\prime}$.
We will also study the D-brane probes in the {\it classical} limit,
and in doing so will self-consistently find results
consistent with the string
coupling remaining bounded throughout the tachyon decay process.
It would be an interesting, but distinct, question to relax
the $g_s\to 0$ limit we consider here and analyze
the quantum dynamics on D-brane probes in these backgrounds,
a question that could be considered both
before and after the tachyons condense.

The classical
world-volume theories of D-branes probing orbifold singularities were
worked out in a beautiful paper by Douglas and Moore~\DouglasSW.
The orbifold group $\Gamma$ has both a geometric action $\hat R$ and an action
$\gamma_R$ on the Chan-Paton indices,
\eqn\Ract{
|\psi,i,j\rangle \ \to\
\gamma^{\vphantom 1}_{R\,ii'}|\hat R\psi,i',j'\rangle
\gamma_{R\,j'j}^{-1}\ .
}
For branes that are free to move away from the orbifold
singularity, there must be a distinct image for each element of $\Gamma$ and so
the Chan-Paton indices transform in the regular representation.
These branes have integer tensions and charges.
Irregular representations correspond to fractional branes bound to the fixed
locus.  We will be interested in the regular case; as we have noted in the
introduction, fractional branes are confined once the singularity is
resolved, but the full mechanism is not understood.

We will consider a D$p$-brane probe that is extended in the directions $\mu
= 0,1,\ldots,p$ and localized in the directions $m = p+1,\ldots,7$ and in the
orbifolded 8-9 plane.  The treatment will be uniform for the IIA or IIB
theories, and for all
$p$ in the respective theories.  We take a single copy of the regular
representation, but the discussion readily extends to $N$ copies.
For $\Gamma =\IZ_n$, $R$ cyclically permutes the D-brane images and so
$\gamma_{R\, jk} = \delta_{j+1,k}$.  The indices $j,k$ are understood to be
defined mod $n$, so in particular
$\gamma_{R\,n1} = 1$.  It is more convenient to work in a basis in which the
spacetime action is not so evident but the spectrum and its quiver
representation are simple:
\eqn\gammaR{
\gamma_{R\, jk} = e^{2\pi i j/n}\delta_{jk}\ .
}

The low energy theory is itself an ``orbifold" of the ${\cal N}=4$
world-volume theory of a D-brane in flat space, obtained by
projecting out gauge theory fields that are not invariant under the
action~\Ract.  The massless open string fields are the vector potential
$A_{\mu jk}$, the collective coordinates $X^m_{jk}$ and $Z_{jk} = (X^8 + i
X^9)_{jk}$, and the spinor $\xi_{jk}$ in the ${\bf 8}$ of $SO(7,1)$ and with
$J_{89} = +\frac{1}{2}$.  The real and imaginary parts of $\xi$ form the
${\bf 16}$ of $SO(9,1)$; we will suppress the $SO(7,1)$ spinor index.  The
orbifold projection \Ract\ on the operation \goodrot, \gammaR\ retains fields
with
$j - k + (n+1) J_{89} = 0$.  The surviving fields are then
\eqn\fields{
A_{\mu\, jj}\ ,\quad X^m_{jj}\ ,\quad Z_{j,j+1}\ ,\quad \xi_{j,j-l}\ ,
}
where $j$ runs from 1 to $n = 2l+1$ and indices are defined mod $n$.  The
conjugates are $\OL Z_{j+1,j}$ and $\OL\xi_{j,j+l}$.

Thus the gauge group is $U(1)^n$, with the collective coordinate $Z_{j,j+1}$
having charge $+1$ under $U(1)_j$ and charge $-1$ under $U(1)_{j+1}$, while
$\xi_{j,j-l}$ has charge $+1$ under $U(1)_j$ and charge $-1$ under
$U(1)_{j-l}$.  The spectrum can be succinctly expressed through ``quiver''
diagrams \DouglasSW.
For each factor in the product gauge group, the diagram has a
node; for $\Gamma = \IZ_n$ these are in one-to-one correspondence with the
range of the Chan-Paton indices.  A field with charge $+1$ under $U(1)_j$ and
charge $-1$ under $U(1)_k$ is denoted by an arrow from node $k$ to node $j$.
For more general representations of $\Gamma$ the gauge group is a product
$\prod_j U(N_j)$ and the arrows represent bifundamentals.
Arrows beginning and ending on the same node are adjoints, which of course
are neutral in the case of $U(1)^n$.  For the example $\Gamma=\IZ_5$, figure~2
shows the separate quiver diagrams for the various fields.
\bigskip

\epsfbox{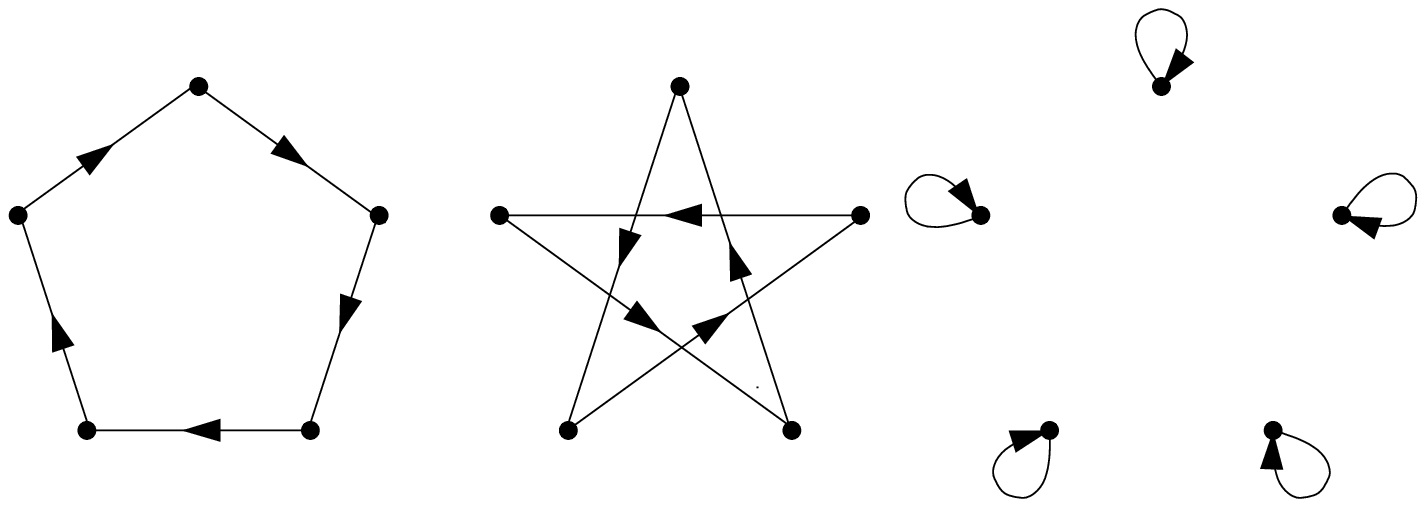}
\smallskip\noindent
Figure 2: Quiver diagrams for the $\IC/\IZ_{5}$ orbifold: for $Z$,
for $\xi$, and for $A_\mu$ and $X^m$.
\bigskip
\noindent

Note that the quiver theory spectrum is invariant under cyclic permutation
of the gauge groups; call this symmetry $\Gamma_Q$.  All gauge invariant
operators inherited from the parent theory, such as $\sum_j F^2_{\mu\nu
jj}$, are scalars under $\Gamma_Q$.  Gauge invariant operators not
descending from gauge invariant operators in the parent theory, such as
$F^2_{\mu\nu 11}$, are not
$\Gamma_Q$ scalars.  Since the lagrangian itself descends from a gauge
invariant operator in the parent theory, it is a scalar, and so $\Gamma_Q$ is
truly a symmetry of the system.
This symmetry is just the realization in the
quiver theory of the orbifold quantum symmetry.  In particular, bulk
twisted modes couple to gauge-twisted operators (\ie\ not $\Gamma_Q$ scalars)
on the brane only in
$\Gamma_Q$ invariant combinations --- a very useful fact in fleshing out the
AdS/CFT dictionary, for example.
We will see that quiver diagrams are a very effective tool for following
the behavior of the probe theory as the singularity decays.

%Of course, once the quivers are in hand the spectrum is trivial to read
%off, something we will exploit shamelessly in the following. Indeed, there
%is a remarkable ammount of physics encoded in the quiver!  It has been
%shown that all orbifold quiver theories are anomaly free, low-fat and more
%fun than a barrel of monkeys.

The potential for the scalars is classically, at the orbifold point,
\eqn\potent{
V = \frac{1}{2} \sum_{j,m} (X_{jj}^m - X_{j+1,j+1}^m)^2 |Z_{j,j+1}|^2
+ \frac{1}{2} \sum_j \Bigl( |Z_{j,j+1}|^2 - |Z_{j-1,j}|^2 \Bigr)^2\ ,
}
where the overall normalization will not be important.
We are interested in the Higgs branch, where $X_{jj}^m$ is independent of $j$
and the $Z_{j,j+1}$ are nonzero.  This corresponds to a D-brane probe of the
orbifold geometry.  On the Coulomb branch the $X_{jj}^m$ depend on $j$ and
the $Z_{j,j+1}$ vanish.  This branch corresponds to the probe separating into
fractional D-branes trapped at the singularity, and it disappears in the
deformed geometry.
On the Higgs branch there is a Yukawa interaction
\eqn\yukawa{
L_{\rm Y} = \sum_j \xi_{j,j-l} \xi_{j-l,j+1} \OL Z_{j+1,j}\ .
}
Note that each interaction forms a closed loop on the quiver diagram.

We now consider the geometry of the Higgs branch.  The vanishing of the
potential \potent\ implies that the magnitude $|Z_{j,j+1}|$ is independent of
$j$.  Of the $n$ $U(1)$ symmetries, the diagonal decouples.  The remaining
$n-1$ gauge symmetries can be used to set the phases of the $Z_{j,j+1}$
equal as well, so that the common value $Z_{j,j+1} = Z$ parameterizes the
branch.  The branch is thus two-dimensional, as it should be for the
interpretation of a probe.  The gauge choice leaves unfixed a
$\IZ_n$ gauge symmetry, whose generator is
\eqn\zngauge{
\exp\biggl( -\frac{2\pi i}{n} \sum_j j Q_j \biggr)\ .
}
This identifies $Z \to e^{2\pi i / n} Z$, so the probe moduli space is indeed
the
$\IC/\IZ_n$ spacetime.  For each of the fields \fields\ there is one massless
mode, where the field is independent of $j$.  This is the correct spectrum
for a D-brane probe.

The moduli space metric, as measured by the probe kinetic term, is obtained by
integrating out the higgsed gauge fields.  In a general gauge, the potential
requires that on the moduli space $Z_{j,j+1} = r e^{i\theta_j}$.  The kinetic
terms are then
\eqn\KineticNoFI{\eqalign{
        L_{\rm k} &= \sum^{n}_{j=1} \Bigl|(\p_\mu +iA_{\mu\, jj}
-iA_{\mu\, j+1,j+1} )Z_{j,j+1} \Bigr|^2 \cr
            &= \sum^{n}_{j=1} \Bigl[ (\p r)^2
+ r^2(\p_\mu \theta_j -B_{\mu j})^2\Bigr]\ .\cr
}}
We have defined the relative gauge potentials, $B_{\mu j} =
A_{\mu\, jj} -A_{\mu\, j+1,j+1} $, which tautologically satisfy the constraint
$\sum B_{\mu j} = 0$.  The total $U(1)$ is unbroken and decouples.  Integrating
out the broken gauge fields subject to the constraint gives
\eqn\intout{
B_{\mu j} = \p_\mu (\theta_j - \tilde\theta)\ ,\quad \tilde\theta =
\frac{1}{n} \sum^{n}_{k=1}
\theta_k \ .
}
Inserting this into the kinetic term gives the manifestly gauge invariant
result
\eqn\Kinetictwo{
L_{\rm k} = n\Bigl[ (\p r)^2 + r^2 (\p \tilde\theta)^2 \Bigr]\ .
}
As deduced above, the periodicity of $\tilde\theta$ is $2\pi/n$.
% Isn't it romantic, Autumn Leaves.  Special thanks to "The Santa Barbara
%Triad"
Rescaling to $\theta = n \tilde\theta$, with canonical period $2\pi$, the
kinetic term becomes
\eqn\Kineticthree{
L_{\rm k} = n\, (\p r)^2 + \frac{r^2}{n} (\p\theta)^2 \ ,
}
corresponding to the metric of a flat $\IZ_n$ cone,
\eqn\MetricNoFI{
         ds^2 =  n\, dr^2 + \frac{r^2}{n}d\theta^2\ ,
}
as expected. For future reference note that we can define $\theta$ as
\eqn\thetadef{
\theta = \arg(Z_{n1} Z_{12} \ldots Z_{n-1,n})\ ;
}
the RHS is gauge invariant, so the period of $\theta$ is manifestly $2\pi$.

The gauge bosons that have been integrated out have masses of order $r
/\alpha'$, while excited string states with masses of order
$\alpha'^{-1/2}$ have been ignored.  The result is therefore valid in the
substringy regime~\DouglasYP, $r\ll \alpha'^{1/2}$.  We have also ignored
quantum corrections in the world-volume theory.  This is valid because the
world-volume fluctuations are open string fields, and we have taken
$g_{\rm s} \to 0$ at the beginning --- we have posed the problem in classical
string theory.

There is a closely related context in which world-volume quantum corrections
would be important.  The world-volume theory of the D1-brane provides
a linear sigma model construction analogous to those in \WittenYC\
of the F-string orbifold
CFT \SilversteinNS.  In this
one must let the quantum world-volume theory flow to the IR fixed point.
In the present case we know independently, from the orbifold construction,
that the fixed point action is the free action~\Kinetictwo.

\newsec{Decay of $\IC/\IZ_n$ in the Substring Regime}

\subsec{Generic Tachyon VEVs: Breaking the Quantum Symmetry}

In the initial stage of the instability, the tachyon VEV is small and so the
geometry is modified only in the substringy region near the tip of the cone.
D-brane probes are therefore the effective tool for investigating the
geometry.  The closed string background determines the low energy quantum
field theory on the probe.  This can be obtained directly from a
calculation of the disk amplitude with a tachyon vertex operator plus open
string vertex operators, as in the appendix of ref.~\DouglasSW.
For our purposes, however, it will suffice to identify the world-volume
theory by matching with the quantum numbers of the closed string
tachyons.

>From the discussion in \S2.1, the tachyons generically break the quantum
symmetry completely, so this will be broken in the world-volume theory.
We are in the substringy regime, so we are interested in the leading effects in
powers of $Z$.  In the potential, this would be a mass term
\eqn\zmass{
\Delta V = \sum_{j=1}^n m_j^2 |Z_{j,j+1}|^2\ .
}
A term of definite quantum charge $k$ would have
a coefficient proportional to $e^{2\pi ijk/n}$.
Since there are tachyons with all charges {except} for the
untwisted $k=0$, one obtains arbitrary masses subject to the constraint
$\sum_j m_j^2 = 0$.  It is then useful to reexpress the mass term as
\eqn\dterm{
\Delta V = -\sum_{j=1}^n \lambda_j
\Bigl( |Z_{j,j+1}|^2 - |Z_{j-1,j}|^2 \Bigr)\ ,
\quad \sum_{j=1}^n \lambda_j = 0\ .
}
The notation is suggested by the supersymmetric case,
where $\lambda_j$ would be
the Fayet-Iliopoulos (FI) coefficient for $U(1)_j$.

On the moduli space we now have
\eqn\diffz{
|Z_{j,j+1}|^2 - |Z_{j-1,j}|^2 = \lambda_j\ .
}
For generic $\lambda_j$ the $Z_{j,j+1}$ are therefore distinct, and one of
these has magnitude less than the rest, say $Z_{12}$.  When this vanishes the
remaining
$n-1$ $Z_{j,j+1}$ are still nonzero.
It follows that $U(1)^n$ is broken to $U(1)$ everywhere on the moduli space,
and so there is no orbifold point.  The moduli space is smoothed;
topologically it is $\IR^2$.  The gauge-invariant combination $Z_{n1} Z_{12}
\ldots Z_{n-1,n}$, which now vanishes linearly when $Z_{12} \to 0$, is a good
coordinate.

We can confirm these conclusions by finding the probe metric.  Define $\rho_j$
iteratively,
\eqn\rhosq{
\rho_j^2 = \rho_{j-1}^2 + \lambda_j\ ,\quad \rho_1 = 0\ .
}
Then with $Z_{j,j+1} = r_j e^{i\theta_j}$, eq.~\diffz\ implies
\eqn\zmod{
r_j^2 = r^2 + \rho_j^2\ , \quad r \equiv r_1\ .
}
The kinetic term is now
\eqn\KineticDef{
        L_{\rm k} = \sum^{n}_{j=1} \Bigl[ (\p r_j)^2
+ r_j^2(\p_\mu \theta_j -B_{\mu j})^2\Bigr]\ .
}
Enforcing the constraint $\sum_j B_{\mu j} = 0$ with a Lagrange
multiplier $\lambda_\mu$, the equation of motion for
$B_{\mu j}$ is
$r_j^2 (\p_\mu \theta_j - B_{\mu j}) = \lambda_\mu$.
Inserting this into the constraint determines the multiplier,
\eqn\getmult{
\lambda_\mu \sum_{j=1}^n \frac{1}{r_j^2} = \p_\mu \theta\ ,
}
where $\theta = \sum_j \theta_j$ is defined as in eq.~\thetadef\ and so has
period $2\pi$.  The action then takes the simple form
\eqn\kineticfinal{
        L_{\rm k} = n(r) (\p r)^2 + \frac{r^2}{n(r)} (\p \theta)^2\ ,
}
where
\eqn\nofr{
n(r) = \sum_{j=1}^n \frac{r^2}{r^2 + \rho_j^2}\ .
}
The corresponding metric is
\eqn\smoothmet{
n(r) dr^2 + \frac{r^2}{n(r)} d\theta^2\ ;
}
for constant $n(r)$ this is the metric \MetricNoFI of a
cone cone of deficit angle $2\pi/n$.  The function $n(r)$ interpolates smoothly
from $n(0) = 1$ (the term
$j=1$) to $n(\infty) = n$.  Thus the metric~\smoothmet\ is nonsingular at the
origin and connects smoothly onto the original $\IC/\IZ_n$ geometry
asymptotically, as in figure~3.

\bigskip
\epsfbox{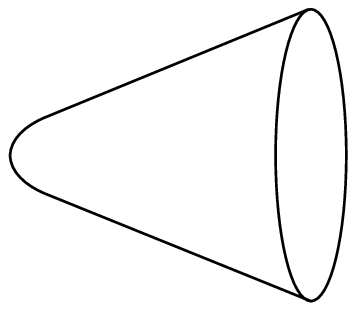}
\smallskip\noindent
Figure 3: $\IC/\IZ_n$ singularity with a twisted tachyon VEV, as seen by a
D-brane probe.
\bigskip

This smoothed geometry differs somewhat from what will eventually emerge in
the gravity regime, depicted in figure~1.  The base of the cone is rounded
rather than flat.  Also, the dilaton is constant: a nontrivial dilaton would
lift the moduli space, through the dependence of the DBI term.  We will see
that in the gravity regime a dilaton must be present, so evidently this is a
higher-order effect.

The exact physical meaning of the D-brane probe calculation here is
a bit indirect.  D-brane probes can observe substringy geometry only on times
long compared to the string scale~\DouglasYP\LiRP, while the decay process
that we are probing takes place on the string time scale.  There are at least
two contexts where the calculation above has a precise meaning.  First, we
could consider a tachyon background which is constant in time and oscillatory
in space, where the wavelength is then long compared to the substringy
geometry. Second, at large $n$ some tachyon masses-squared are of order $1/n$.
Even when neither of these contexts is relevant, we expect that the
qualitative conclusion about the geometry is correct, and this is all that we
will need.

Again, our analysis of the gauge theory is entirely classical.
The non-supersymmetric gauge theories do not look unstable in this
approximation at the orbifold point.  The tachyon instability is a closed
string tree effect and so a one-loop open string effect.
In the context of AdS/CFT duality, we would expect to see this instability in
the gauge theory; we will return to this point in \S4.

Finally, note that the resolved geometry is topologically trivial.  Thus,
unlike the supersymmetric ALE singularity, there is no interpretation in
terms of collapsing cycles at real codimension two.
However, in \S3.3, and in \S5 where we consider
the case of real codimension four orbifold singularities, we
will see many parallels with the supersymmetric case.

\subsec{World-sheet Linear Sigma Model}

As we noted above, the D1-brane gauge theory provides
the starting point for a LSM
representation of the F-string world-sheet theory.
Let us digress slightly to explain the picture of the tachyon
decay process which emerges from this point of view.

The LSM description involves considering a simple gauge
theory in the UV which flows to the world-sheet CFT of interest
in the IR \WittenYC.  In the context of quiver theories on D1-branes at
orbifold points, the classical moduli space is the orbifold
space (as we reviewed in \S2.2), which is the target
space for the F-string world-sheet CFT.  Based on this and the
discrete symmetries of the theory arising for appropriate choice of
theta angles, it was argued in \SilversteinNS\ that the
D1-brane quiver theory provides a linear sigma model
formulation of the orbifold CFT, with the caveat that without supersymmetry
one must fine-tune away the quantum potential on the moduli space
in order to reach the orbifold CFT in the IR
(which then enjoys an accidental
supersymmetry).

We are interested in the effect on the world-sheet CFT when the
tachyon VEVs are turned on in spacetime, which means in terms
of a renormalization group analysis that a relevant
operator is added to the world-sheet CFT action (taking the tachyons
at zero spacetime momentum).  We would like to describe
this deformation from the UV LSM quiver theory.
As we have discussed, the tachyons
transform under the quantum symmetry in the IR CFT, and this
symmetry exists already in the UV quiver theory.  Therefore
we can identify twisted operators in the UV theory which will generically
mix with the twisted-sector
tachyon vertex operators in the IR.  The twisted couplings
of interest include the $\lambda_j$ in \dterm\ above.  These
are the most relevant twisted deformations in the UV, and we
will focus on their effects.

The RG flow diagram of this theory appears as in the following
figure.

\bigskip
\epsfbox{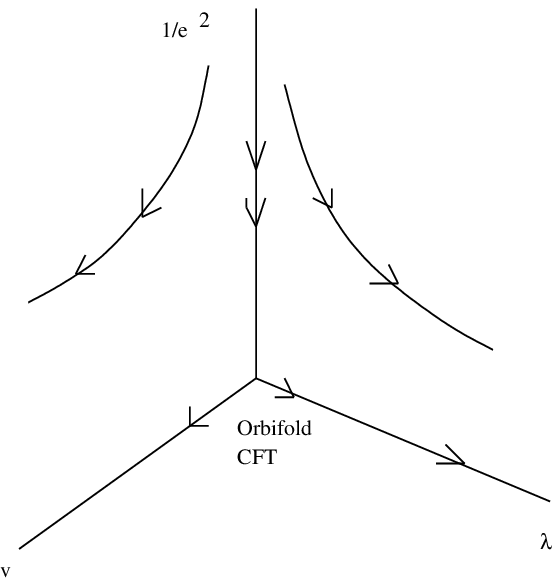}
\smallskip\noindent
Figure 3.5: Flow diagram for the linear sigma model.
\smallskip

We consider flow toward the IR, keeping track
of the indicated couplings $e$ (the gauge coupling)
and $\lambda$, and on a third axis
the relevant couplings $v$ in the scalar potential of the theory which
drive the flow away from the desired IR world-sheet theory; these
last we tune away as discussed in \SilversteinNS.
The flow proceeds toward stronger gauge coupling $e$.
As we turn on $\lambda$, the vacuum manifold of the LSM smoothes out,
as we discussed above.  For large $\lambda$, integrating out the
massive degrees of freedom in the LSM
we obtain a nonlinear sigma model whose RG flow proceeds
toward infinite flat space, as we will see in \S4.
For small $\lambda$, as we flow toward the IR we expect
generically for $\lambda$ to mix with the tachyon vertex
operators, which are relevant operators so that the flow
proceeds away from the orbifold CFT fixed point.

Putting this together, the simplest joining of the two regimes leads again
to a picture where the tachyon VEV induces flow from
the orbifold CFT to smooth flat space.

\subsec{Special Tachyon VEVs: Annealing the Quiver}

We have considered a generic tachyon VEV, which in the quiver theory breaks
all $U(1)$s and resolves the singularity completely.  It is interesting to
consider instead partial resolutions of the singularity.  Depending
on the choice of twisted deformation we turn on,
we will find that such resolutions can lead to quasi-orbifolds, which
have no free world-sheet CFT description, or to real orbifolds, which
do.  We will start with an example of the former case and then proceed to the
transitions between real orbifolds that are our main interest.

Consider for
example the case that
$\lambda_1 = - \lambda_2 > 0$, for which eq.~\diffz\ implies that one
bifundamental is greater than the rest,
\eqn\zstep{
|Z_{12}|^2 = |Z_{j,j+1}|^2 + \lambda_1\ ,\quad j \neq 1\ .
}
The maximum unbroken gauge symmetry is now $U(1)^{n-1}$, where all
$Z_{j,j+1}$ other than $Z_{12}$ vanish, so we expect
that the symmetry is partially resolved to $\IZ_{n-1}$.

The theory near the fixed point can be elegantly described in terms of {\it
annealed} quiver diagrams.
As an explicit example, let us analyze the $\IC/\IZ_5$
orbifold, whose quiver diagrams were given in figure~2.  Figure~4 shows the
first step in the annealing.  In the neighborhood of the fixed point, the
bifundamental $Z_{12}$ has a relatively large VEV and
breaks $U(1)_1 \times U(1)_2$ to the diagonal $U(1)$.  Thus, in the second
line of figure~4 we have collapsed nodes 1 and 2.

%
%    **FIG Z5 TO Z4 QUIVER ANNEALING**.
%
\smallskip
\epsfbox{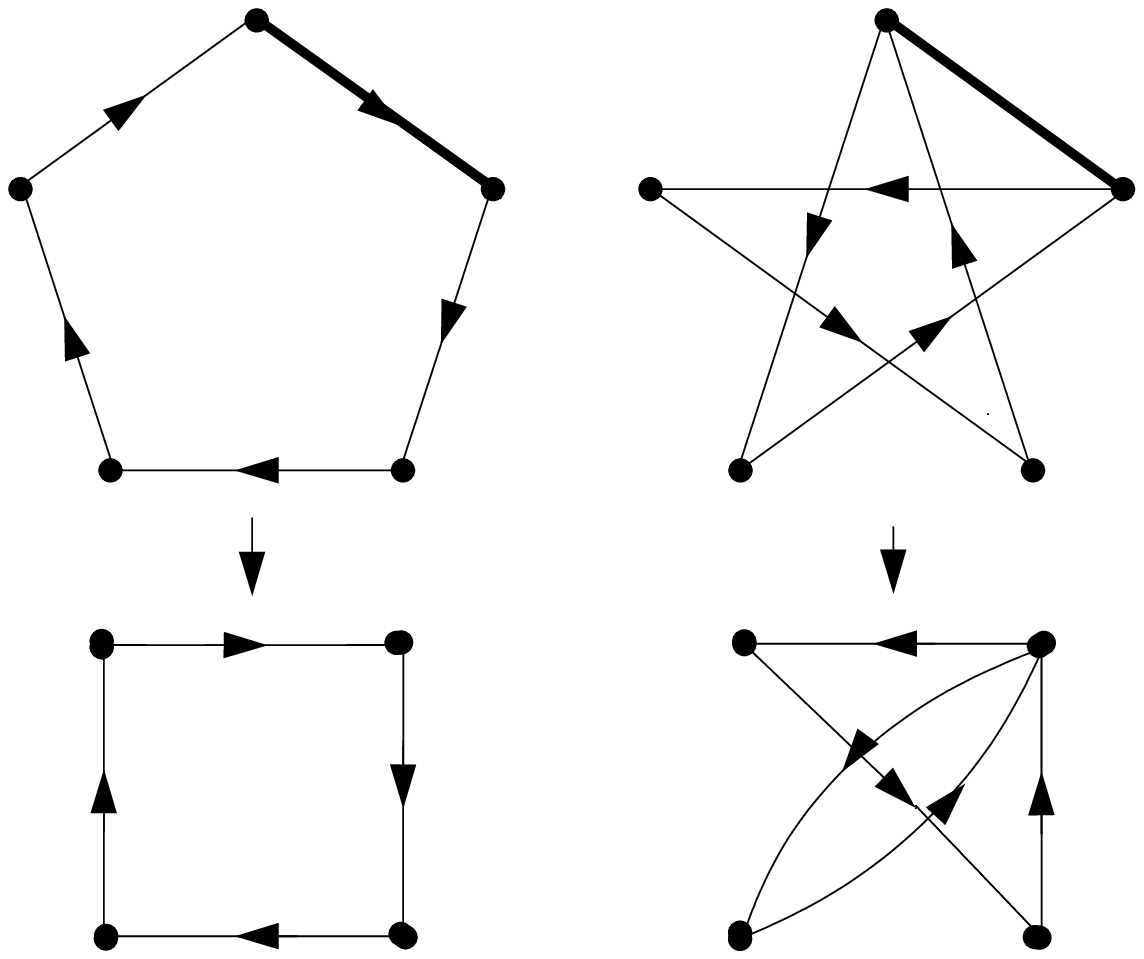}
\smallskip
\noindent
Figure 4: Partially annealed $\IC/\IZ_{5}$ scalar and fermion quivers.  The
scalar
$Z_{12}$ is indicated in bold.  In the low energy theory the
nodes 1 and 2 are identified.
\smallskip

The scalar $Z_{12}$ decouples from the low energy theory, its
magnitude fixed by the potential and its phase absorbed by higgsing; thus it
is omitted from the annealed diagram.  The adjoint scalar $X_{11}^m -
X_{22}^m$ accompanying the broken $U(1)$ is also lifted by the potential.
Finally, the mass term $\xi_{14} \xi_{42} \OL Z_{21}$ removes two fermions,
so the final quiver diagram is shown in figure~5.

\smallskip
\epsfbox{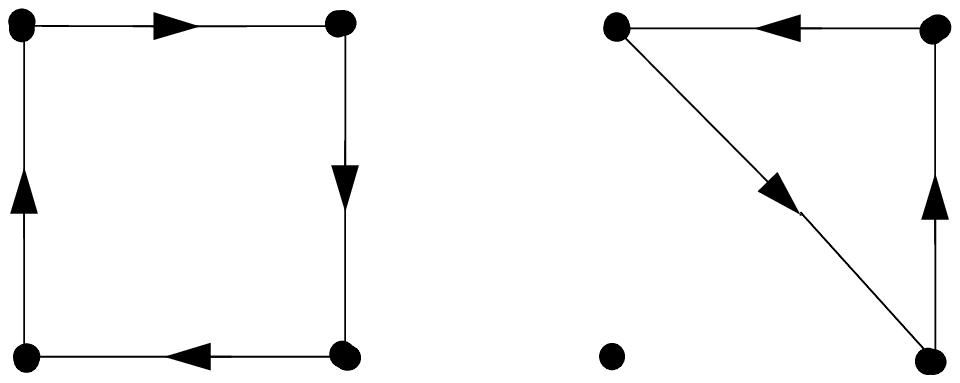}
\smallskip
\noindent
Figure 5: Final annealed $\IC/\IZ_{5}$ scalar and fermion quivers.
\smallskip

The scalar spectrum is the same as for a $\IC/\IZ_{4}$ orbifold in bosonic
string theory, and the metric~\nofr\ seen by a D-brane probe has a $\IZ_{4}$
singularity. The geometry is as in figure~6, with a $\IZ_4$ singularity in a
space whose asymptotic geometry is $\IC/\IZ_{5}$.

\bigskip
\epsfbox{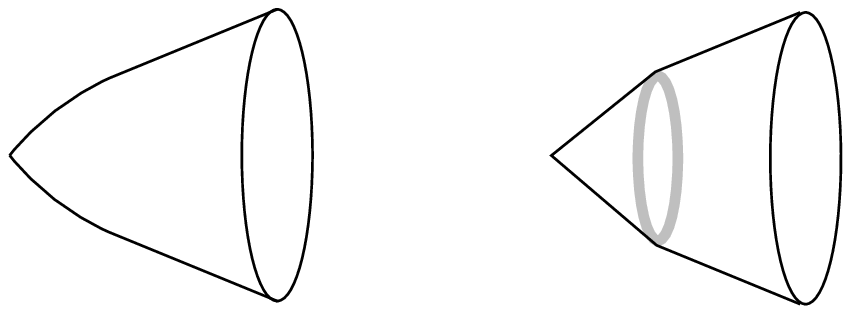}
\smallskip
\noindent
Figure 6: Asymptotic $\IC/\IZ_{n}$ geometry with a $\IC/\IZ_{n'}$
singularity, with $n' < n$, as seen in the substringy and gravity
regimes.
\smallskip

The fermion spectrum is not of quiver form.  This is not surprising,
as we know that there is no orbifold construction of the
supersymmetric type II string on the $\IC/\IZ_{4}$
singularity.  Rather, this must be a quasi-orbifold, not based on a free CFT,
as discussed in \S2.  However, by turning on additional
Fayet-Iliopoulos terms, and so a second scalar VEV, we can flow to the $\IZ_3$
quiver as shown in figure~7; it is easy to check that the Yukawa terms lift
no additional fermions.

%    **FIG TWO SCALAR VEVS**,
\bigskip
%\epsfbox{z5_quiver.anealed.good.eps}
\epsfbox{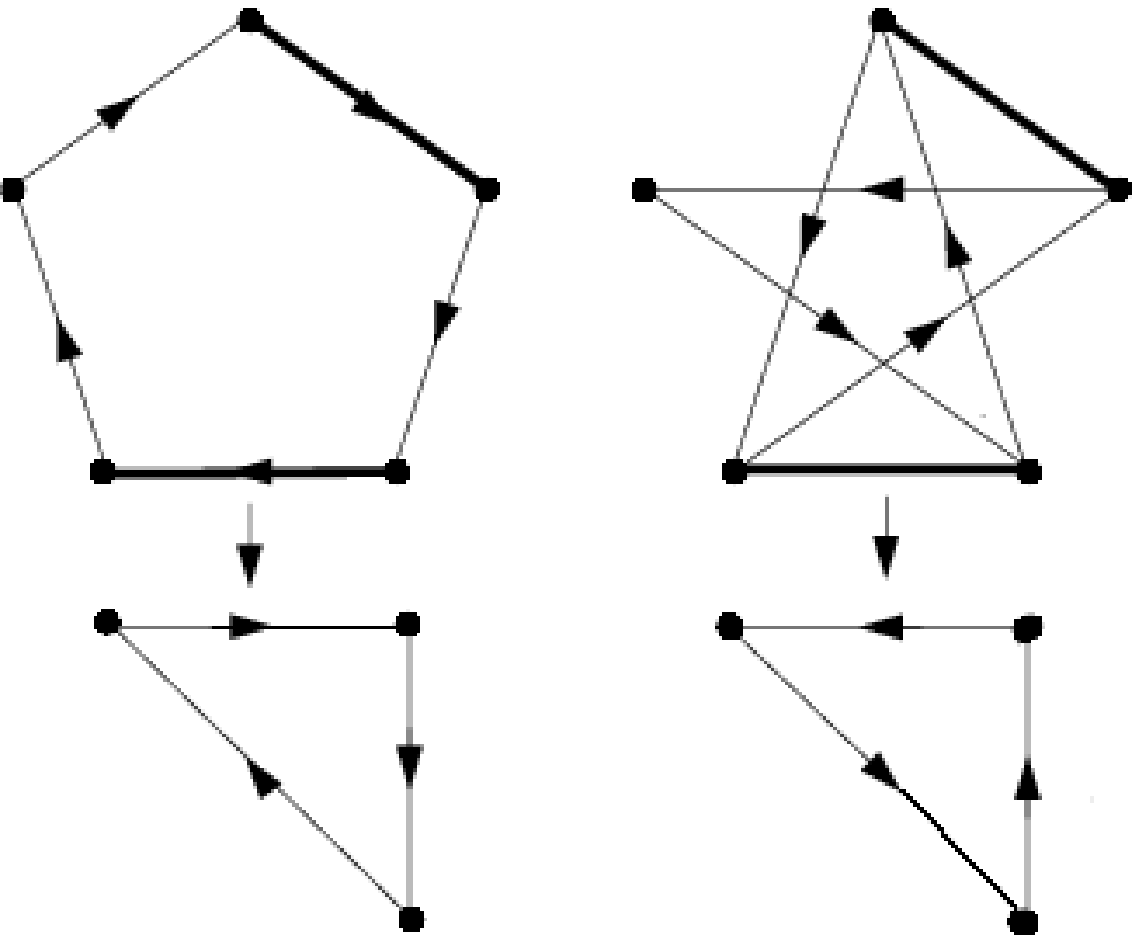}
\smallskip
\noindent
Figure 7: The massless sector of $\IC/\IZ_{5}$ with two scalars
turned on gives the $\IC/\IZ_{3}$ quiver!
\smallskip

More generally, the $\IC/\IZ_{2l+1}$ singularity can decay to
the $\IC/\IZ_{2l-1}$ singularity, if the FI terms are such that $Z_{12}$ and
$Z_{l+1,l+2}$ decouple.  (It can also flow to a variety of quasi-orbifold
singularities.)  For the true orbifold case, the quiver diagram has an obvious
$\IZ_{2l-1}$ symmetry.  This is not a subgroup of $\IZ_{2l+1}$, but emerges
as an accidental symmetry (in the technical sense) of the low-energy theory.
This process can be repeated until we reach the trivial $\IC/\IZ_1$ orbifold,
without tachyons.

%    **FIG Z1 QUIVER**.
\bigskip
\epsfbox{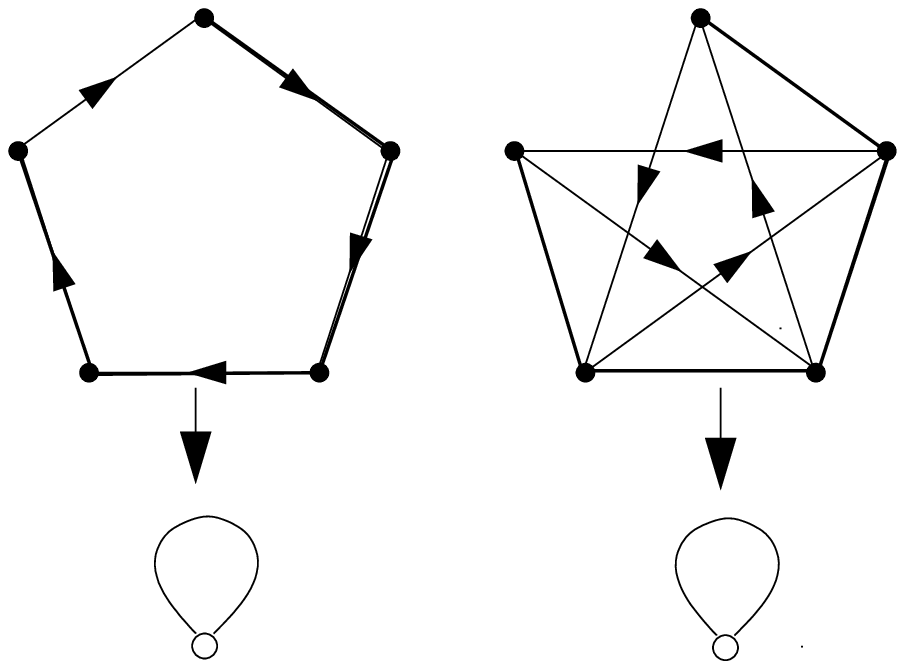}
\smallskip
\centerline{Figure 8: Decay to the $\IC/\IZ_{1}$ space.}
\smallskip

\noindent
The spectrum is simply a free chiral supermultiplet, with SUSY reappearing
as an accidental symmetry under successive quiver annealings.  The order of
liftings does not particularly matter.  As long as all but one scalar receive
generic VEVs the result
is inevitably the quiver for SUSY flat space, regardless of the geometries or
effective quivers at intermediate scales.  Since the bifundamentals couple
to relative gauge potentials, this maximally higgses the system; we cannot
lift all the scalars without changing the number of degrees of freedom in our
theory.
In complex codimension two, the story will be much richer, as there is an
infinite family of SUSY quivers to which a generic tachyonic quiver can
decay.

Note that the decays to lower-order singularities require specific FI terms of
no particular quantum symmetry, so they arise from a linear combination of
different tachyons.  Since the tachyons have different lifetimes, the
singular point will not be a static configuration.  Presumably it is possible
to fine-tune the initial conditions so that the geometry develops the lower
order singularity as it enters the gravity regime, where it will remain
on top of its tachyon potential.

\newsec{Decay of $\IC/\IZ_n$ in the Gravity Regime}

In the preceding section we found that the initial effect of the tachyonic
singularity is to smooth the geometry.  As with any tachyon, an essential
question is the nature of the final state: does the tachyon potential have a
minimum, or does the instability continue without end?  The D-brane probe
analysis in the previous section breaks down when the size of the smoothed
region reaches the string scale.  We do not have tools to probe this regime,
so will study the question by going beyond it to the regime of small curvature.
If we were to find that the RG flow in that regime carries us back to higher
curvature, this would indicate the presence of a minimum with curvature of
order the string scale.  In fact, we will find that the flow goes toward
ever-smaller curvature.\foot{We cannot exclude the possibility of a fixed
point with curvature of order the string scale, but the fact that both the
substring and gravity geometries evolve toward smaller curvature strongly
suggest that this flow continues smoothly through the stringy regime.}
Thus the geometry evolves forever, generating
an arbitrarily large region of arbitrarily small curvature, which contains a
lower-order singularity if the initial state has been appropriately fine-tuned.

\subsec{RG Flow}

We now study the RG flow of the world-sheet NLSM corresponding to a background
of the massless closed string fields.  Owing to discrete symmetries, we need
only consider the metric and dilaton.  Note that there is no explicit tachyon
field in this regime. The instability, whose initial stage is represented by a
tachyon in the orbifold description, would now be a property of the solutions
to the low energy field equations. The RG equations are
%driven by the
%string beta functions, which are beautifully and lovingly presented in
%Joe's fantastic book, which i use almost every day, as:
\eqn\rgflow{\eqalign{
\dot G_{MN} &= - \beta[G_{MN}] + \nabla_{M} \xi_N
+ \nabla_{N}\xi_M\ ,\cr
\dot \Phi &= - \beta[\Phi] + \xi^M \nabla_M \Phi\ ,
}}
where $G_{MN}$ is the string metric, a dot denotes the logarithmic
derivative with respect to world-sheet length scale $\ell \partial_\ell$, and
\eqn\BetaFns{\eqalign{
 \beta[G_{MN}] &= \alpha' R_{MN} + 2\alpha' \nabla_{M} \nabla_{N}
\Phi\ , \cr
 \beta[\Phi] &= \alpha' (\nabla \Phi)^2 - \frac{\alpha'}{2} \nabla^2\Phi\ . \cr
}}
The vector field $\xi_M$ is arbitrary and represents the freedom to make a
spacetime coordinate change with the change of world-sheet scale.  A convenient
choice is $\xi_M = \alpha' \nabla_M \Phi$, so that
\eqn\simpflow{
\dot G_{MN} = - \alpha' R_{MN}\ ,\quad
\dot \Phi = \frac{\alpha'}{2} \nabla^2\Phi\ .
}
In these coordinates the flow of the metric does not depend on the dilaton;
this is possible because the dilaton does not appear in the flat world-sheet
action.

The perturbation leaves a $(7+1)$-dimensional free field theory, so the problem
is essentially two dimensional. It is convenient to work in conformal gauge,
because the flow	\simpflow\ preserves that gauge.  Thus,
\eqn\confgauge{
ds^2 = e^{2\omega} (d\rho^2 + \rho^2 d\theta^2)\ ,
}
where for generality we consider an arbitrary periodicity $0 \leq \theta \leq
2\pi/\nu$.  In this gauge, the metric~\MetricNoFI\ for a cone of opening angle
$2\pi/n$ corresponds to
\eqn\flatpsi{
\omega = \biggl( \frac{\nu}{n} - 1 \biggr) \ln \rho + \hbox{constant}\ .
}
In conformal gauge the RG is
\eqn\confRG{
\dot\omega = \frac{\alpha'}{2} e^{-2\omega} \hat\nabla^2 \omega
\ ,
}
where $\hat\nabla^2$ is the Laplacian for the flat metric  $d\rho^2 + \rho^2
d\theta^2$, which is $\partial_\rho^2 + \rho^{-1} \partial_\rho$ for a
cylindrically symmetric solution.

Let us analyze this first for the transition $n \to n-2$ at large $n$, where
the change in the metric is small.  The geometry is as depicted in figure~6,
with a $\IC/\IZ_{n-2}$ cone at the origin, going smoothly to a $\IC/\IZ_{n}$
cone at large radius. In coordinates with $\nu = n-2$, the boundary conditions
are
\eqn\rgbc{
\omega(\rho \to 0) = \hbox{finite}\ ;\quad \omega(\rho \to \infty)
 \to -\frac{2}{n} \ln
\rho \ . }
We can then linearize, $\dot\omega = \alpha' \hat\nabla^2 \omega/2$.  A simple
solution, obtained by the Fourier transform on the covering space, is
\eqn\linsol{
\omega(\rho,\ell) = -\frac{1}{n} \biggl( \ln\ln(\ell/\ell_0) +  \int_0^{u_0}
\frac{du}{u} (1 - e^{-u}) \biggr) \to\ -\frac{1}{n} \int_0^{u_0}
\frac{du}{u} (1 - e^{-u})
\ , }
where $u_0 = \rho^2 / 2 \alpha' \ln(\ell/\ell_0)$; in the second form the
$\ln\ln$ term has been conveniently absorbed in an
$\ell$-dependent rescaling of
$\rho$. The solution depends only on $u_0$, so the radius $\rho_{\rm t}$ of the
transition region grows with increasing world-sheet length scale,
$\rho_{\rm t} \sim \ln(\ell/\ell_0)^{1/2}$.  Pointwise in the IR the system
approaches a $\IC/\IZ_{n-2}$ cone everywhere.  Asymptotically, all solutions to
the diffusion equation with the given boundary conditions will have the same
form. The dilaton satisfies a diffusion equation as well and any initial
dilaton gradient will similarly diffuse outward.

For the full nonlinear evolution~\confRG\ we do not have a simple analytic
result, but given the diffusive nature of the equation we expect that in
general the smoothed area depicted in figure~3 grows without bound.  Hence our
conclusion that the flow found in the substring region, toward smaller
curvature, continues indefinitely in the gravity region.

There are two reasons that one might doubt this result.  The first is the
Zamolodchikov $c$-theorem, showing irreversibility of the
flow of the central charge~\ZamolodchikovGT.  Here we start with an orbifold CFT
of canonical central charge (15 in all for the type II string).  In the IR, we
claim that the theory flows pointwise to flat spacetime, again with canonical
central charge.  The reason that this is consistent is that the noncompactness
of the target space invalidates the $c$-theorem~\PolchinskiDY.  There are other
cases of CFT theorems that are invalid in noncompact target spaces.  The classic
example is the holomorphicity of conserved currents, which does not hold for
the world-sheet currents associated with rotational invariance in noncompact
directions~\BanksYZ.  For the $c$-theorem, the basic objects are the vacuum
expectation values of operator products.  The string world-sheet vacuum fills
out the entire target manifold, a familiar IR effect, so the region of
curvature makes a contribution of measure zero.

A second reason that one might have expected the opposite result is the
example of compact spaces of positive curvature, which flow to greater
curvature.  We claim that the difference of boundary conditions in the compact
and noncompact cases accounts for the differing behaviors.  In fact, there is
a simple monotonicity result that makes this clear.  From the differential
equation~\simpflow\ it follows that
\eqn\volflow{
\ell \partial_\ell \int d^2x \sqrt{G} = -\frac{\alpha'}{2} \int d^2x
\sqrt{G}R\ .
}
For a manifold of spherical topology, the RHS is $-4\pi\alpha'$ and so the
volume is monotonically decreasing.  The curvature must at some point become
stringy, and the low energy theory break down.  For the noncompact manifold
the integral is not defined, but one can consider the integral interior to a
circle of some given radius (over a region such as depicted in figure~3).  The
smoothing of the singularity does have the effect of reducing this volume,
whereas flow back toward a singular cone would increase the volume in
contradiction to the flow~\volflow.

Finally, we might also be interested in the case that the original singularity
is part of a compact space.  Most simply, consider $T^2/\IZ_3$, which
is a flat space of spherical topology, with three $\IC/\IZ_3$ singularities
each of deficit angle $4\pi/3$.  From the $c$-theorem, or from eq.~\volflow,
one concludes that the space eventually flows to large curvature.  The three
singularities begin to smooth, until the smoothed regions merge to form a
rough sphere, which then evolves toward smaller radius.

\subsec{Dynamical Evolution}

We now consider on-shell evolution,
\eqn\onshell{
\beta[G_{MN}] = \beta[\Phi] = 0\ ,
}
with the same $\beta$-functions~\BetaFns.  This is now a three-dimensional
problem, since the solution depends on time.  It is convenient to work in the
Einstein frame, where this system is just a
massless scalar canonically coupled to the metric.  The initial metric is
again assumed
to interpolate from $\IC/\IZ_n$ at infinity to $\IC/\IZ_{n'}$ at the origin,
with $n' < n$.  This is true in both the Einstein and the string frames, because
we assume that the dilaton is nonsingular at the origin, while it goes to a
constant at infinity (where the evolution has not yet reached).

We do not have analytic solutions for this problem, but it is easy to deduce
the general form of the solutions.
The constraint equations require that the change in deficit angle be
accompanied by energy density of matter.  Since we can solve
the equations with the NS three-form
field strength set to zero, so that the only matter involved is the
massless dilaton, this energy
must be dilaton gradient and kinetic energy.  This
dilaton field will radiate outward at the speed of light, as in figure~1.  The
time scale of the initial decay, before the gravity regime, is the string
scale, so this sets the initial width of the dilaton pulse and the kink in the
geometry, which then gradually broadens due to dispersion.  For $n' = n-2$ at
large $n$, an analytic treatment is again simple.  The dilaton satisfies a
massless wave equation in flat spacetime, and the backreaction on the metric
is a perturbative effect.

As a check, let us look for static solutions in the gravity
regime, which would have corresponded to minima of the tachyon potential that
are visible in this regime.  We will
take the most general form with $SO(7,1) \times SO(2)$ spacetime
symmetry:
\eqn\TachLineElement{
 ds^2=e^{2\sigma(r)}\eta_{\mu\nu}dx^\mu dx^\nu +e^{2f(r)}(dr^2 +r^2 d\theta^2) \
. }
This is slightly more general than elsewhere, in that we allow the $(7,1)$
directions to be warped; note also a slight change of notation, $\mu,\nu = 0,
\ldots, 7$.  The dilaton field equation is
\eqn\StaticDilEQ{
\frac{\Phi''}{\Phi'} - 2\Phi' = -\frac{1}{r} - 8\s'
\quad \Rightarrow\quad (e^{-2\Phi})'=\frac{c_1}{r}e^{-8\s}
}
with integration parameter $c_1$.
The $\mu\nu$ curvature equation reads
\eqn\EaaEOM{
\frac{(r\s')'}{r\s'} +8\s' = 2\Phi'
\quad \Rightarrow\quad e^{2\Phi} = c_2 r\s'e^{8\s}
}
with integration parameter $c_2$.
Putting these together gives $\Phi' =  -c_1 c_2 \s'/2$,
and so
\eqn\StaticPhiSolution{
e^\Phi \propto (\ln r/r_0)^{-c_1 c_2 /2(8+c_1c_2)}\ ,\quad
e^\s \propto (\ln r/r_0)^{1/(8+c_1c_2)}\ .
}
These are doubly unacceptable: they do not go over to the unperturbed behavior
at large $r$, and they have a singularity at finite $r = r_0$.  Only the flat
cone, the exceptional solution with $\Phi$ and $\s$ constant, survives.

% sleeeeeeeeeeppppppp.....................
It is interesting again to consider the $T^2/\IZ_3$ orbifold, with
nonsupersymmetric singularities in a compact space.  We cannot follow the
behavior analytically, but might expect that after the dilaton pulses have
begun to cross the compact space, the time-averaged behavior will be that of
a positively curved radiation dominated spacetime.  Thus, it will reach a Big
Crunch in finite time, beyond which we cannot follow the evolution.  One
supposition would be that the compact dimensions effectively disappear, leaving
an eight-dimensional noncritical string theory \deAlwisPR.
However, the simplest
background in that theory --- the linear dilaton --- has the
wrong symmetries to
be the endstate of our evolution, as the dilaton gradient is spacelike.

\subsec{Application to AdS/CFT}

In ref.~\KachruYS\ it was argued that orbifolding should commute with AdS/CFT
duality, so that
the dual of the orbifolded gauge theory is IIB string theory on
the orbifolded spacetime.  This expectation is based on the fact that
a duality like the AdS/CFT correspondence concerns a single system with
two dual descriptions; if orbifolding makes sense on one side of
the duality then the procedure can be mapped to the equivalent dual
description of the system given a complete duality dictionary
translating between them.  In the absence of supersymmetry,
if the orbifolding procedure produces a consistent physical system, this
requires any instabilities that arise to match up between the two
equivalent descriptions.  The leading instability that arises
in a string background is that of interest here, namely the tachyons.  In
freely-acting orbifolds on the sphere component of the $AdS_p\times S^q$
geometry, the spectrum is classically tachyon-free. Non-freely acting
orbifolds on the $S^q$ do have tachyons, and this has been considered at
small 't Hooft parameter \AdamsJB, where the gauge theory is perturbative.  Now
let us consider the situation at large 't Hooft parameter, where the AdS
description is good.

The AdS description starts with $N$ coincident D3-branes extended in the 0123
directions. The orbifold produces a (7+1)-dimensional fixed plane.  This
plane contains the D3-branes and extends in four transverse directions.
The AdS curvature is small on the string scale and so locally on the fixed
plane the initial instability is the same as in flat spacetime.  In
particular, the decay will release a given energy per unit volume of the fixed
plane, as measured in a local inertial frame.  The invariant volume element is
$(r/R_{\rm AdS})^{3}  d^3x\, (r/R_{\rm AdS})^{-4} r^3 dr$, where $x$
coordinatizes the field theory dimensions and $r$ is a coordinate along the
radial direction of $AdS_5\times S^5$, with metric $(r/R_{\rm AdS})^2 dx^2
+(r/R_{\rm AdS})^{-2} (dr^2+r^2d\Omega^2)$.  The translation to the global
conserved energy brings in an additional factor of $r/R_{\rm AdS}$, so the
total energy released per unit gauge theory volume is simply
\eqn\diverg{
\int_0^\infty dr\,r^3 \sim \Lambda^4\ .
}
That is, it diverges quarticly in the gauge theory.

We can make a simple model of how this divergence might arise in the gauge
theory.  Consider a state a $U(1)$ gauge theory where we add a $+$ and a $-$
charge in a volume of linear size $\ell$.  The kinetic energy is of order
$2\ell^{-1}$, but this is reduced somewhat by the gauge theory potential, for a
net $\{ 2 - O(g^2) \}\ell^{-1}$.  Extrapolation would suggest a possible
instability at large $g^2$ (to be precise, this theory will have a Landau pole
in the UV, so we must imagine a cutoff).  In a globally supersymmetric theory,
positivity of the energy is guaranteed and so this instability is absent;
thus, supersymmetric field theories can make sense at large coupling.
However, for nonsupersymmetric theories there is no guarantee that they make
sense at strong coupling.  Indeed the result \diverg\
suggests an instability of
just this sort: in a conformal theory we can produce pairs on any scale $\ell$,
and the integral over all scales produces a quarticly divergent result.  Note
that this is much more severe than the instabilities normally encountered in
field theories (such as symmetry breaking), which are IR effects and release a
finite energy per unit volume.  It is difficult to see how this instability
could have any sensible final state.

Indeed, the AdS picture is similarly pathological.  We can quantitatively
study the large-$n$ case $\IZ_n \to \IZ_{n-2}$, because the dilaton
essentially satisfies a free wave equation on the $AdS_5 \times S^5$ covering
space,
\eqn\waveq{
\frac{R_{\rm AdS}^2}{r^2} \partial_t^2 \Phi = \frac{r^2}{R_{\rm AdS}^2}
\partial_\perp^2 \Phi\ .
}
The orbifold breaks the $SO(6)$ symmetry of $S^5$ so the dilaton is a
superposition of different angular states.  For angular momentum $L$,
\eqn\lapl{
\partial_\perp^2 = \partial_r^2 + \frac{5}{r} \partial_r - \frac{L(L+4)}{r^2}\
.}
Imagine that the decay starts everywhere at once at $t=0$.  This condition is
conformally invariant so the dilaton is a function only of $rt$.  The wave
equation~\waveq\ then becomes an ordinary differential equation for
$\Phi_L(rt)$, and $rt = R_{\rm AdS}^2$ is a singular point.  From the dominant
terms near the singular point one finds that
\eqn\phil{
\Phi_L \sim (R_{\rm AdS}^2 - rt)^{-3/2}
}
for every partial wave $L$.  Thus, the energy density diverges at finite time
for any $r$; this occurs when a geodesic from $(r,t) = (\infty,0)$ reaches a
given radius, carrying the information about the divergent energy release at
large radius.

Again, this instability is a property of large 't Hooft parameter, and is
not inconsistent with the much milder instability found at small 't Hooft
parameter in ref.~\AdamsJB.  Note that we have assumed that the 't Hooft
parameter does not run, as holds at large $N$.  If the full $\beta$ function
were in fact asymptotically free, then the theory would be stable in the UV,
and the instability that we are discussing would set in only below some
scale.  In this event it is possible that there would be a stable final state.

%\subsec{A Night in Tunisia}

\newsec{$\IC^2/\IZ_n$ Orbifolds and Non-SUSY to SUSY Flows}

\subsec{Orbifolds and Quivers}

One of the interesting results of the study of open string
tachyons has been the possibility of realizing stable branes,
in particular SUSY branes, by open string tachyon condensation
\Opentach \BerkovitsHF.
In this section, we study closed string tachyon
condensation on $\IC^2/\IZ_n$ orbifolds by generalizing the D-brane probe
approach of \S3\ to this case.
We will exhibit various transitions from non-supersymmetric, tachyonic
$\IC^2/\IZ_n$ orbifolds to supersymmetric ALE spaces, and provide
an infinite sequence of such flows which allows us to realize
any SUSY ALE space via closed-string tachyon condensation (or
more generally a combination of marginal deformation and tachyon condensation).

The discussion will parallel the $\IC/\IZ_n$ case.
All the orbifolds that we consider will be based on a twist of the form
\eqn\zn{
R =
\exp\biggl\{ \frac{2\pi i}{n} (J_{67}+ kJ_{89}) \biggr\}\ ,
 }
depending on two integers $n$ and $k\ (\hbox{mod}\ 2n)$.  We will denote the
group generated by $R$ as $\IZ_{n(k)}$.  On spinors with $J_{67}$
and $J_{89}$ charge
$s_{67} = s_{89} = \pm \frac{1}{2}$, $R$ acts as $e^{\pm 2\pi i(k+1)/2n}$.
On spinors with $-s_{67} = s_{89} = \pm \frac{1}{2}$ it acts as
$e^{\pm 2\pi i(k-1)/2n}$.
The condition that $R^n = 1$ on spinors forces $k$ to be odd.  If $k$ is
$\pm 1$, then $R$ leaves half of the $D=10$ spinors invariant and so produces
the familiar supersymmetric $A_{n-1}$ orbifold (for reviews see
\AnselmiSM\JohnsonCH).  For other values of $k$, at least some of the twisted
sector ground states are tachyonic.  If $2n$ is divisible by $k+1$ or
by $k-1$, then $R^{2n/(k+1)}$ or $R^{2n/(k-1)}$ leaves some spinors
invariant.  The associated twisted sector ground state is
massless, and indeed is the same as the corresponding twisted sector state in
the supersymmetric orbifold (but note that the respective cases $k+1 = 2$ and
$k-1 = 2$ are trivial).

Let us now consider a D-brane probe in this background.
Define
\eqn\zs{
Z^1 = X^6 + iX^7\ ,\quad  Z^2 = X^8 + iX^9\ .
}
For the world-volume spinor in the ${\bf 16}$ of $SO(9,1)$, its component with
$(s_{67}, s_{89}) = (- \frac{1}{2},- \frac{1}{2})$ will be denoted $\chi$ and it
component with $(s_{67}, s_{89}) = (- \frac{1}{2},+ \frac{1}{2})$ will be
denoted $\eta$ (the remaining two components are the conjugates).  The $SO(5,1)$
spinor indices, respectively ${\bf 4}$ and ${\bf
4'}$, are suppressed.
Using the techniques discussed in \DouglasSW\ and \S2,
one finds the world-volume theory to be a $U(1)^{n}$ quiver theory with matter
content
\eqn\ctwofields{
A_{\mu\, jj}\ ,\quad X^m_{jj}\ ,\quad Z^1_{j,j+1}\ ,\quad Z^2_{j,j+k}\ ,
\quad \chi_{j,j-q-1}\ ,\quad \eta_{j,j+q}\quad (k \equiv 2q+1)\ .
}

The classical scalar potential is
\eqn\potone{
V = {\rm Tr}\biggl\{ \frac{1}{2} [Z^1,\OL Z^1]^2 + \frac{1}{2}
[Z^2,\OL Z^2]^2 + \big|[Z^1,Z^2]\big|^2 + \big|[Z^1,\OL Z^2]\big|^2
\biggr\}\ . }
Using the Jacobi identity this can also be rewritten
\eqn\pottwo{\eqalign{
V &= {\rm Tr}\biggl\{ \frac{1}{2} \big([Z^1,\OL Z^1] -
[Z^2,\OL Z^2]\big)^2 + 2\big|[Z^1,\OL Z^2]\big|^2
\biggr\}\cr
&= {\rm Tr}\biggl\{ \frac{1}{2} \big([Z^1,\OL Z^1] +
[Z^2,\OL Z^2]\big)^2 + 2\big|[Z^1,Z^2]\big|^2
\biggr\}\ . }}
The Yukawa terms are
\eqn\yukone{
L_{\rm Y} = {\rm Tr} \Bigl\{ [Z^1, \chi]\, \eta + [Z^2, \chi]\, \OL\eta +
\hbox{h.c.} \Bigr\}\ .
}

\subsec{The Example $\IZ_{2l(2l-1)}$: Non-SUSY $\IZ_{2l}$ to
SUSY $\IZ_2$}

Now we analyze the case $k = n-1$, where $n = 2l$ must be even
because $k$ is odd.  Here
\eqn\rcaseone{
R =  \exp\{ {2\pi i} J_{89} \}\exp\{ {2\pi i} (J_{67} -
J_{89})/2l \}
}
is the same as in the supersymmetric case except for a factor of
$\exp\{ {2\pi i} J_{89} \} = (-1)^{\bf F}$, which breaks supersymmetry.
Note however that the special case $l=1$, the $\IC^2/{\IZ}_{2(1)}$
orbifold, is supersymmetric:
$R = \exp\{ {2\pi i} (J_{67} + J_{89})/2\}$ leaves invariant spinors such that
$s_{67} = - s_{89}$.

Before exciting tachyons, the geometry is the same as for the
supersymmetric orbifold.  In particular, on the probe moduli
space the condition that $V$ vanish gives
\eqn\ctwomod{
Z^1_{j,j+1} = Z^1\ ,\quad Z^2_{j+1,j} = Z^2\quad
\hbox{(independent of
$j$)}}
up to gauge transformation.
Thus the probe has two complex moduli, as it should.  The origin, where the
$U(1)^{2l}$ gauge symmetry is restored, is a $\IZ_{2l}$ singularity as in
\S2.2.

Before discussing the generic decay, it is interesting to consider first
deformations that preserve a $\IZ_2 \subset \IZ_{2l}$ quantum symmetry.  This
$\IZ_2$ acts on the $l^{\rm th}$ twisted sector as $(-1)^l$, so only states
twisted by powers of $R^2$ can have VEVs.  Since $R^2 = \exp\{ {2\pi i} (J_{67}
- J_{89})/l \}$, these sectors are exactly the same
as for the {\it supersymmetric} $\IC^2/\IZ_{l(-1)}$ orbifold.  In particular,
there are no tachyons, so we are actually considering marginal deformations.
The perturbation of the gauge theory is then a supersymmetric $D$-term
\eqn\ctwozmass{
\Delta V = -\sum_{j=1}^{2l} \lambda_j D_j\ ,\quad
D_j = |Z^1_{j,j+1}|^2 - |Z^2_{j+1,j}|^2 -
|Z^1_{j-1,j}|^2 + |Z^2_{j,j-1}|^2 \ ,
}
where the sign of each term is determined by the $U(1)_j$ charge
(note that for a $k=-1$ orbifold $Z^1$ and $Z^2$ are in chiral superfields,
while for $k=+1$ it would be $Z^1$ and $\OL Z^2$). An overall additive
constant is ignored. As in \S2.2, $\sum_{j=1}^{2l} \lambda_j = 0$, while the
$\IZ_2$ quantum symmetry requires that $\lambda_j = \lambda_{j+l}$.  Now
consider the deformed moduli space; focus on the second of forms \pottwo\ and
note that the first term there is just $\sum_{j=1}^{2l} D_j^2$.  The vanishing
of the second term requires that
\eqn\zprod{Z^1_{j,j+1} Z^2_{j+1,j} \equiv \alpha
}
be independent of $j$.
Minimizing the $D$-terms then sets $D_j =
\lambda_j$, which determines all of the
magnitudes in terms of $|Z^1_{12}|$ and $\alpha$.  Finally, the phases can
be gauged away except for
$\sum_{j=1}^{2l} \arg Z^1_{j,j+1}$, giving four real moduli in all.

There are still singularities.  Consider the subspace $\alpha = 0$.  The
condition $D_j = \lambda_j$ determines
\eqn\diffmag{
|Z^1_{j,j+1}|^2 - |Z^2_{j+1,j}|^2 = \rho_j + x\ ,
}
where $\rho_j = \rho_{j-1} + \lambda_j$ and $x$ is undetermined.  When
$x = -\rho_{j_0}$ for some ${j_0}$, both $Z^1_{{j_0},{j_0}+1}$ and
$Z^2_{{j_0}+1,{j_0}}$ vanish. Further, the $\IZ_2$ quantum symmetry implies that
$Z^1_{{j_0}+l,{j_0}+l+1}$ and
$Z^2_{{j_0}+l+1,{j_0}+l}$ vanish as well.  There are then two
unbroken $U(1)$'s, namely $\sum_{j={j_0}+1}^{{j_0}+l} Q_{j}$ and
$\sum_{j={j_0} + l+1}^{{j_0}} Q_j$, where $Q_j$ is the $U(1)_j$ charge and
$j$ is defined mod
$2l$.  Thus, these $l$ points of restored gauge symmetry, which are
generically distinct, are
$\IZ_2$ singularities on the moduli space.

Thus far the discussion is the same as for the resolution of a {\it
supersymmetric} $\IC^2/\IZ_{2l(-1)}$ singularity while preserving a $\IZ_2$
quantum symmetry: the result there would be a $\IZ_{2(-1)}$ orbifold of a smooth
$\IZ_l$ ALE space. The difference for us is that the final orbifold operation
here contains an extra factor of $(-1)^{\bf F}$, so it must be ${\IZ}_{2(1)}$.
Naively one might expect this orbifold point to be nonsupersymmetric, but the
discussion at the beginning of this subsection shows that it is supersymmetric
with the opposite supersymmetry from that respected by $R^2$.  One can think of
the final picture as follows: we resolve the $\IC^2/\IZ_{l(-1)}$ orbifold
generated by
$R^2$ into a smooth ALE space preserving half of the supersymmetry, and then
make a
${\IZ}_{2(1)}$ orbifold which locally would preserve the other half.  In other
words, we have a space of
$SU(2)_1 \subset SO(4) = SU(2)_1 \times SU(2)_2$ holonomy, with $l$ orbifold
singularities whose holonomy is in $SU(2)_2$.
The space as a whole has no
supersymmetry, but half of the supersymmetry survives in the smooth region and
the other half locally at the orbifold points.  In the limit that the marginal
deformation is taken to infinity, we simply have a supersymmetric
$\IC^2/{\IZ}_{2(1)}$ space, without tachyons.

We can verify this by examining the quivers.  At the orbifold point, the
potential for the vanishing fields $Z^1_{{j_0},{j_0}+1}$,
$Z^2_{{j_0}+1,{j_0}}$, $Z^1_{{j_0}+l,{j_0}+l+1}$ and
$Z^2_{{j_0}+l+1,{j_0}+l}$ is quartic so they
are massless, while all other scalars are massed up.
One unbroken $U(1)$ acts
on indices $j = j_0,\, j_0+l+1$, and the other on indices $j = j_0+1,\, j_0+l$.
Expanding the Yukawa coupling~\yukone\ in components, one finds that
$\eta_{j_0+1,j_0+l}$ and
$\eta_{j_0+l+1,j_0}$ do not appear in terms with scalar expectation values, so
these remain massless (note that these are neutral under the unbroken
$U(1)$'s).
There must therefore also be two massless linear combinations of
$\chi$'s; these come in the bifundamental representation of
the unbroken $U(1)^2$.
%closer inspection shows that these are given by $\chi_{j,j+l}$ being
%constant on the range $j_0 + 1 \leq j \leq j_0 + l$ and on the range
%$j_0+l+1
%\leq j
%\leq j_0+2l$.
The correlation between $U(1)$ charges and $SO(5,1)$ quantum
numbers is the same as for the $\IC/{\IZ}_{2(1)}$ orbifold, namely the spectrum
\ctwofields\ at $k=1$ with $\eta$ in the adjoint and $\chi$ in
the bifundamental representation of the gauge group.

We now turn to the generic twisted state background.  The full D-brane probe
analysis is less useful here, for two reasons.  The first is that without any
connection to supersymmetry, the quantum symmetry alone does not fix the form of
the quadratic mass terms (specifically, the ratio of $Z^1$ and $Z^2$
masses); it requires the
calculation of a disk amplitude, as in the appendix to ref.~\DouglasSW.
More critically, for general mass terms allowed by the quantum symmetry, there
is no probe moduli space.  This is not a problem --- from the spacetime point of
view it is the same effect that a dilaton background would have --- but it
makes it difficult to give a geometric interpretation in the substring regime.

Fortunately, we can largely deduce the fate of the instability by expanding
around the deformation already considered.  Let us first deform
$\IC^2/\IZ_{2l(2l-1)}$ along directions that preserve the $\IZ_2$ quantum
symmetry as above, so as to have an orbifold of $SU(2)_2$ holonomy in a space of
$SU(2)_1$ holonomy.  The orbifold locally is supersymmetric and so has marginal
deformations in the twisted sector.  These correspond to blowing the orbifold
points up into smooth $\IZ_2$ ALE spaces of $SU(2)_2$ holonomy.  Thus we have
small patches of $SU(2)_2$ holonomy in a larger region of $SU(2)_1$ holonomy.
This is only an approximate solution to the equations of motion, and will in
time evolve to a space of generic holonomy and expand indefinitely
as in the $\IC/\IZ_n$ case.

Note that the second blowing-up will not be exactly marginal, as the coupling
to the $SU(2)$ curvature will break supersymmetry and presumably drive the
marginal direction to be tachyonic.  If the extent of initial blowing-up is
reduced, so as to condense the two steps towards one, the ${\IZ}_{2(1)}$
twisted state will become more tachyonic, so we seem to connect smoothly onto
the original string-scale tachyon.

There is a seeming paradox here, whose resolution provides an elegant check on
our picture.  The initial $\IC^2/{\IZ}_{2l(2l-1)}$ orbifold is an exact 
CFT, and
so its tree-level energy (as measured by the $1/r^2$ falloff of the metric) is
zero.  There is a tree-level tachyon, and so the final state should have
negative energy when the kinetic energy of the outgoing pulse is
subtracted.\foot
{This paradox did not arise for $\IC/\IZ_n$, because in two dimensions a conic
deficit angle {\it is} an ADM energy.}
Does this not violate a positive energy theorem?  In fact, there
is no such theorem: negative energy configurations of asymptotic ALE geometry
exist \lebrun.\foot
{We would like to thank G. Horowitz for informing us about these spaces and
explaining their significance, as well as sharing insights from
his investigations into GR solutions for the $\IC^2/\IZ_n$ cases \horowitz.}
There is a negative energy theorem for any geometry that admits spinor fields
going to a constant at infinity~\WittenMF\GibbonsGU.  The geometries of
ref.~\lebrun\ admit spinors, so it must be that any smooth spinor field is {\it
antiperiodic} under the asymptotic ALE identification.  This is
precisely the geometry of the ${\IZ}_{2l(2l-1)}$ orbifold.

In the above example and the others we will consider in
this section, we have studied in detail the substring regime
using D-brane probes, and in the case of marginal deformations, 
we have also studied the regime far away from the original orbifold
point using inheritance from a related SUSY orbifold.  
Once tachyons turn on and the system evolves
into the gravity regime, we have not analyzed 
the subsequent GR solutions as explicitly as in the
$\IC/\IZ_n$ case.  However, the following indicates that the behavior is as
before.
Consider a configuration of negative energy.  If the size of the
configuration  is scaled up by a factor $\lambda$, the energy scales as
$\lambda^2$ ($\lambda^4$ from the volume and $\lambda^{-2}$ from the
derivatives).  This implies that the potential is unbounded below in this
direction.

\subsec{The Example $\IC^2/\IZ_{2l(3)}$:  Non-SUSY $\IZ_{2l}$ to SUSY $\IZ_l$}

These results have a resemblance to phenomena that
have been observed in open string systems.  The existence of a tachyon,
which disappears as one goes along a marginal direction, is the same as in a
D-brane/anti-D-brane system, where the string-scale tachyon at small
separation goes over to a long-range attraction as the branes are separated.
The decay of a nonsupersymmetric configuration to a
supersymmetric configuration plus outgoing radiation is also familiar.

There are many other similar flow patterns that can be
deduced by studying the quiver theories as we have done
for the above case.  One interesting sequence is
for $n = 2l$ and $k=3$,
\eqn\zn{
R =
\exp\biggl\{ \frac{2\pi i}{2l} (J_{67}+ 3J_{89}) \biggr\}\ ,
 }
whose quiver diagrams are shown in figure~9.
\smallskip
\epsfbox{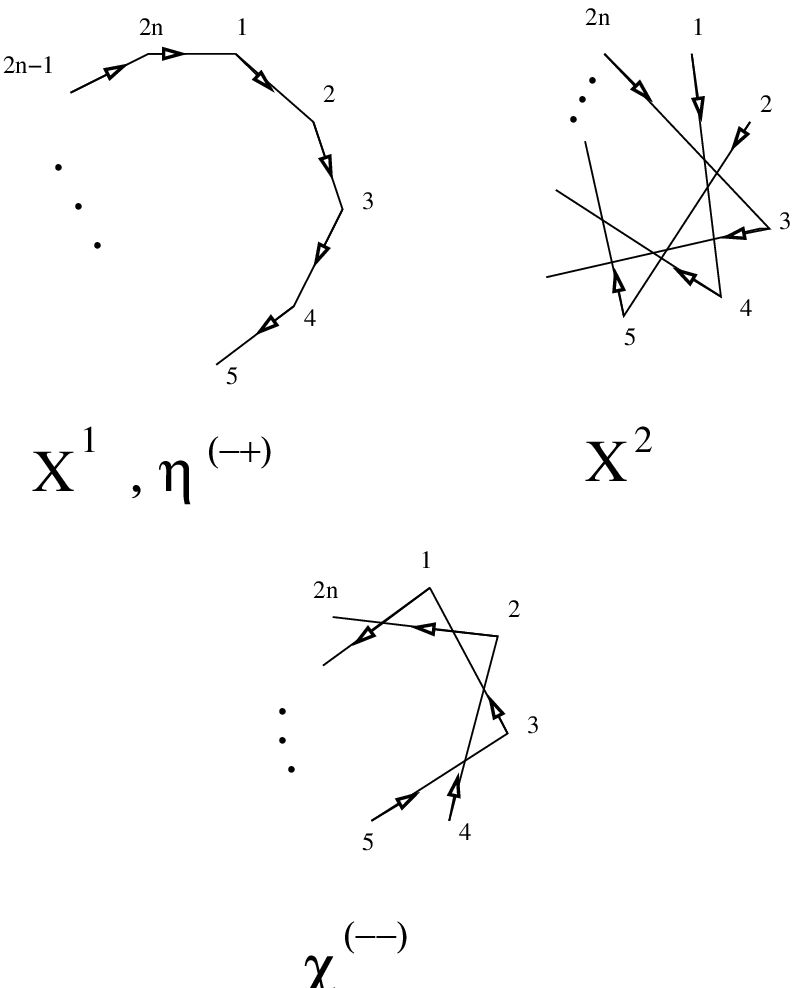}
\smallskip
\noindent{Figure 9: $\IC^2/\IZ_{2l(3)}$ quivers}
\bigskip
\noindent
In this case $R^l = \exp\{ i \pi (J_{67} - J_{89}) \}$ is the
same as for the supersymmetric $\IC^2/\IZ_{2(-1)}$ orbifold.  In parallel
with the previous example, we first excite only marginal states from the
sector twisted by $R^l$.  This preserves as $\IZ_l$ subgroup of the original
$\IZ_{2l}$ quantum symmetry.

The Fayet-Iliopoulos terms then satisfy
\eqn\fialt{
\lambda_j = (-1)^{j+1} \lambda
}
where we take $\lambda > 0$.  The quantum symmetry requires that
$Z^1_{2p-1,2p}$, $Z^1_{2p,2p+1}$, $Z^2_{2p-1,2p+2}$, and $Z^2_{2p,2p+3}$
be independent of $p$, and the
$D$-terms are minimized when
\eqn\extwovevs{
|Z^1_{2p-1,2p}|^2 + |Z^2_{2p-1,2p+2}|^2 =
|Z^1_{2p,2p+1}|^2 + |Z^2_{2p,2p+3}|^2 + \lambda\ ,
}
while $Z^1_{2p-1,2p} Z^2_{2p,2p+3} = Z^1_{2p,2p+1} Z^2_{2p-1,2p+2}$.  When
$Z^1_{2p-1,2p} =
\lambda^{1/2}$ with all other VEVs vanishing, a $U(1)^l$ is restored ---
namely $Q_{2p-1} + Q_{2p}$ for all $p$ --- so this is a $\IZ_l$ singularity.
Before taking into account interactions that give mass to some fields, the
quiver diagrams thus collapse to those depicted in Figure 10.
\smallskip
\epsfbox{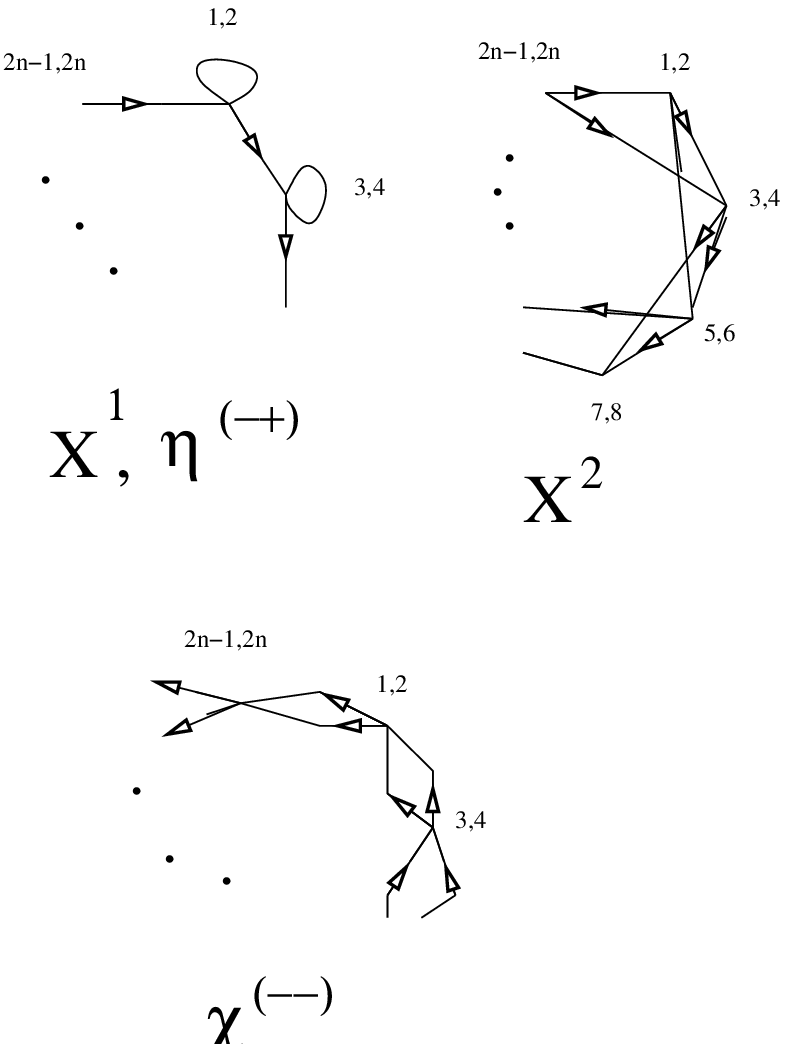}
\smallskip
\noindent{Figure 10: $\IC^2/\IZ_{l}$ quivers from collapse of
$\IC^2/\IZ_{2l(3)}$, including massive fields}
\bigskip

We next must determine which of the fields in figure
10 mass up in the transition.
On the $Z^1$ diagram, the adjoint representations are removed: the potential
fixes the magnitudes and the Higgs mechanism removes the phases, leaving the
result in Figure 11.  On the $Z^2$ diagram, the $|[Z^1,Z^2]|^2$ term gives
masses to $Z^2_{2p, 2p+3}$, so that the components depicted in figure~11
remain massless.  Of the fermions, only half of the
components appear in the mass matrix, namely $\eta_{2p,2p}$ and
$\chi_{2p+1,2p-1} - \chi_{2p, 2p-2}$, leaving the fermions
depicted in figure 11.  In particular, the $\eta$
are in the adjoint representation and
the $\chi$ are in
the $(q, q+1)$ bifundamental.
\smallskip
\epsfbox{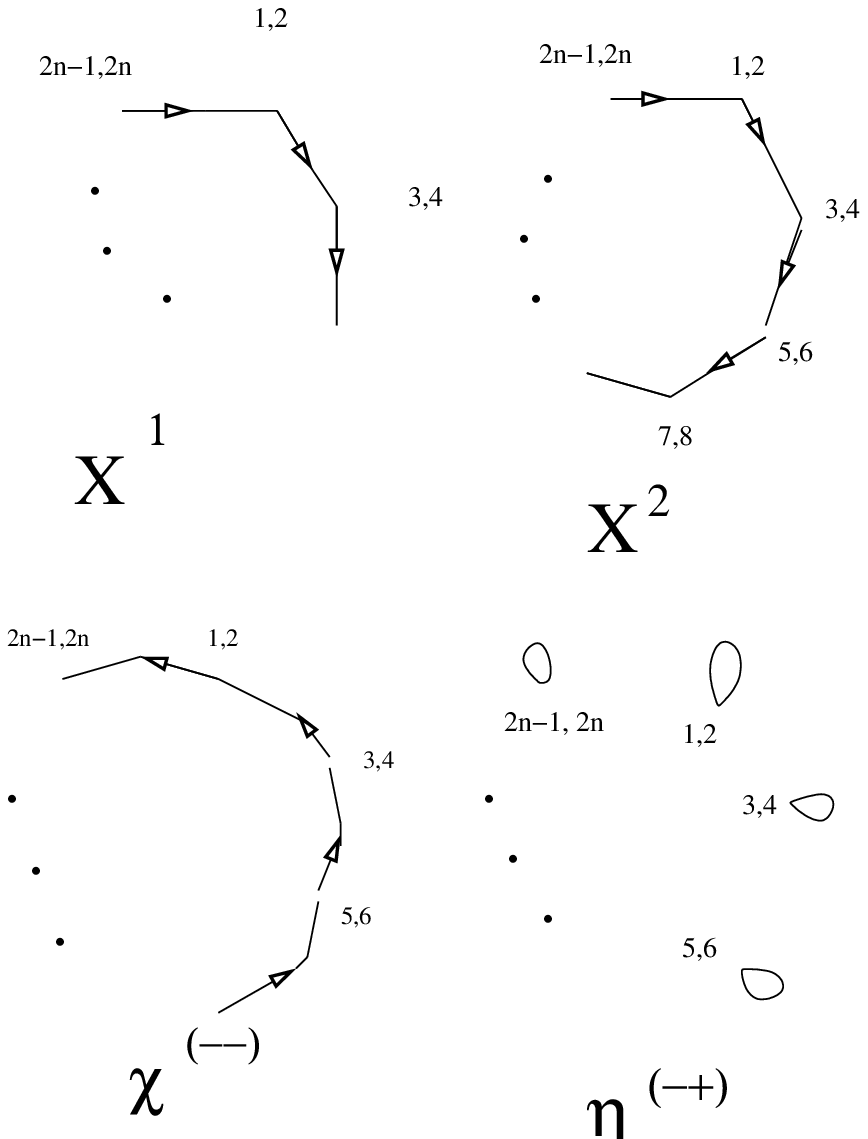}
\smallskip
\noindent{Figure 11: $\IC^2/\IZ_{2l(3)}$ quivers after twisted-state
condensation}
\bigskip

Altogether, we are left in Figure 11 with the quiver
theory corresponding to D-branes at a
$\IC^2/\IZ_{l(1)}$ orbifold point, which is supersymmetric but with the
opposite supersymmetry from the $R^l$ orbifold.  Thus the interpretation is
parallel to the previous example: the marginal direction blows up the
orbifold into a manifolds of smooth $SU(2)_1$ holonomy, which is orbifolded by
$\IZ_{l(1)} \subset SU(2)_2$.

As a check, consider the low energy theory near the fixed point.  We have
$R = \exp(2\pi  i J/2l)$, where $J = J_{67} + 3 J_{89}$.  This operator in the
original theory becomes
\eqn\lerot{
J_{67} + 3 J_{89} + \frac{1}{2}\sum_{p=1}^{l}  (Q_{2p} - Q_{2p-1})\ ,
}
in the low energy theory,
because this is the linear combination including the broken generators that
leaves the background invariant.  This acts on the massless fields as
\eqn\lerottwo{
Z^1_{2p,2p+1} \to 2 Z^1_{2p,2p+1}\ ,\quad   Z^2_{2p-1,2p+2} \to 2
Z^2_{2p-1,2p+2}\,
}
and so it acts as $\tilde J = 2( J_{67} + J_{89} )$ in the low energy theory.
The orbifold operation $\exp(2\pi  i \tilde J/2l)$ is then $\IZ_{l(1)}$ in the
low energy theory.

There is another orbifold point, where
$Z^2_{2p-1,2p+2} = \lambda^{1/2}$ with all other VEVs vanishing.  The
analysis of the previous paragraph shows that this is a $\IZ_{l(-3)}$, which
is nonsupersymmetric for $l > 2$.

In summary, we can obtain all supersymmetric ALE
orbifolds by descent from nonsupersymmetric ones.
The $\IC^2/\IZ_{4(3)}\to \IC^2/\IZ_{2(1)}$ flow is common
to both this sequence and the one discussed in \S5.2.

%%*** $\IZ_{5(3)}$ example --- you may not want to put in as much detail ***

\subsec{The Example $\IC^2/\IZ_{5(3)}\to \IC/\IZ_{2(1)}$:  Tachyon
Condensation}

Since both of the above examples involved marginal as well as
tachyonic deformations, it is interesting to ask whether
there are in fact examples where such transitions between
non-supersymmetric and supersymmetric ALE spaces proceed
exclusively by tachyon condensation, without any marginal
component.  The following simple
example exhibits this possibility (which we expect to
be generic).  We will make the assumption that the twisted
deformations we turn on in the quiver
world-volume QFT can be accessed by adjusting modes in
the tower of twisted states in the closed string sector.
It would be interesting to check this generic assumption
more explicitly as in \DouglasSW.

Start with the orbifold $\IC^2/\IZ_{5(3)}$.  We can choose three independent
$\lambda_j$ such that the D-terms induce VEVs for $Z^1_{45}$,
$Z^1_{51}$, and $Z^1_{23}$.  This preserves a $U(1)^2$ subgroup
of the $U(1)^5$ gauge symmetry, generated by combinations of
charges $Q_4+Q_5+Q_1$ and $Q_2+Q_3$.  Plugging these VEVs into
the component expansion of the interaction terms \pottwo\yukone\
as before, we find that the spectrum reduces to that of the
$\IC/\IZ_{2(1)}$ quiver theory, with gauge group
$U(1)^2$, $\eta$ in the adjoint and $\chi$, $Z^1$, and $Z^2$
transforming as bifundamentals.  This theory does not
have effectively supersymmetric subsectors, in contrast
to those in \S5.2\ and \S5.3.  So given our assumption
about the availability of these deformations in the closed
string spectrum (including those that put the
Lagrangian in supersymmetric form), this provides an example of a truly
tachyonic transition from a non-supersymmetric ALE space
to a supersymmetric one.

\newsec{Dualities, Fluxbranes, and the Type 0 Tachyon}

The results in the preceding sections
describe transitions between different ALE spaces (including
flat space) by processes in which the string coupling remains
bounded.  While this is sufficient for our purposes, it
is also instructive to consider the predictions that
these results imply for processes in dual descriptions of
the system.  In particular we will consider $T$-dual descriptions, in the
angular direction, of the orbifolds that we have considered, as well as the
addition of R-R Wilson lines.  The duals thus involve NS5-branes,
fluxbranes, and the type 0 tachyon.

\subsec{$\IC/\IZ_n$ at Large $n$}

An angular direction along which we rotate in performing a $\IZ_n$
orbifold projection ends up $n$ times smaller than in
the parent theory, so for $n$ large it is of interest
to $T$-dualize along this direction in the region near
the origin.  Near but not at the origin, there is a
subspace that looks approximately like a cylinder, with twisted
strings playing the role of winding modes around the
$S^1$ direction of the cylinder:
\bigskip
\epsfbox{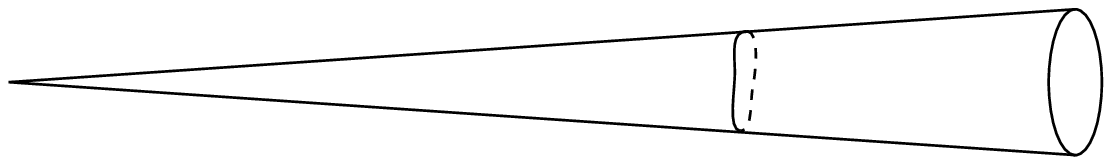}
\smallskip
\noindent
Figure 12: The $\IC/\IZ_{n}$ cone at large $n$, with a twisted closed string.
\smallskip
\noindent
In other words, there are
many low-lying twisted states, which become Kaluza-Klein states in the
$T$-dual description.  The formal $T$-dual of the cone metric \MetricNoFI\ is
\eqn\tdualcone{
ds^2 = n dr^2 + \frac{n \alpha'^2}{r^2} d\tilde\theta^2\ .
}
Also, the dilaton is now position-dependent,
\eqn\tdualdil{
e^{\Phi} = g_{\rm s} \frac{\sqrt{n\alpha'}}{r}\ .
}

In the large-$n$ limit the orbifold operation \goodrot\ is a small
rotation times $(-1)^{\bf F}$, so on the circle that we are $T$-dualing
fields are twisted by $(-1)^{\bf F}$.  Such a twist has three effects on
the $T$-dual description.  First, the bulk theory is twisted by $(-1)^{\bf
F}$, so it is the type 0 theory (type 0B if we began with IIA, and type 0A if
we began with IIB).  Second, in going around the $T$-dual circle there is a
twist by
$(-1)^{\bf Q}$, where $\bf Q$ is the quantum symmetry dual to the twist
$(-1)^{\bf F}$.  That is, type 0 fields that descend from the type II theory
are periodic, while type 0 fields that do not descend (including the type 0
tachyon) are antiperiodic.  Third, the periodicity of the dual coordinate is
halved, $0 \leq \tilde\theta \leq \pi$.

The
$T$-dual description is valid out to $r \sim n\sqrt{\alpha'}$, beyond which
the $T$-dual circle is small and the original circle is large.  It also breaks
down for $r < \sqrt{\alpha'}$, where the curvature becomes large.  Thus, the
apparent divergence of the dilaton \tdualdil\ is irrelevant, as we could have
expected since the orbifold description is manifestly weakly coupled.  We do
not have any good description in this region; it is some sort of effective
`wall' in spacetime, whose properties can be deduced from the exact orbifold
description.  One property of the wall that is not evident in the
metric~\tdualcone\ is the breaking of translation invariance in the
$\tilde\theta$-direction.  The twisted modes of the orbifold
transform under the
finite $\IZ_n$ quantum symmetry rather than the
infinite group $\IZ$ characterizing true winding
modes on a cylinder.  In the $T$-dual description, this
means that the continuous translation symmetry along
the dual angular circle is
broken to a discrete $\IZ_n$ symmetry~\GregoryTE.  This suggests
that the wall is actually a line of $n$ branes (defined
broadly as defects
which break translation invariance) spaced equally
along the $T$-dual circle.  In the case
of $\IC^2/\IZ_n$, this picture is well understood, as we will review
shortly, but for $\IC/\IZ_n$ we do not know of any suitable candidate branes.

The twisted state tachyon of the original theory is just the
bulk type 0 tachyon in the $T$-dual description.
The multiplicity of excited tachyons associated with
the eight-dimensional fixed plane on the orbifold
side maps on the $T$-dual side
to the multiplicity of modes of the ten-dimensional
type 0 tachyon.  Because of the $(-1)^{\bf
Q}$ twist the decay is most rapid at small $r$.
The type 0 tachyon in ten dimensions has
${\alpha^\prime\over 4}m^2= -{1\over 2}$, and
the antiperiodic boundary condition should shift this upward by an amount of
order the inverse radius of the dual circle.  Indeed, the most
tachyonic mode has
\eqn\mosttach{
{\alpha^\prime\over 4}m^2= -{1\over 2}+{1\over{2n}}
}

For the partial
resolution $n \to n-2$, it seems that the wall relaxes into a lower energy
state while a metric and dilaton perturbation (given by the $T$-dual of the
picture in \S4) propagates outward.
For the full decay $n \to 1$, the
tip of the cone and the associated low-lying states disappear at the speed
of light.  In the $T$-dual picture, it seems that the wall, where
our control breaks down, is propagating
to larger $r$  at the speed of light.  At larger $r$, the
angular direction gets smaller in this T-dual picture.
It would be interesting to try to extract from this
a prediction for the type 0 tachyon, but this is not
immediate in our system here because the initial wall is present
to act as a seed for the decay.

\subsec{$\IC^2/\IZ_n$ and NS5-Branes}

For the orbifold $\IC^2/\IZ_n$ at large $n$ and fixed $k$, the angular
direction generated by $J_{67} + k J_{89}$ is again small and a $T$-dual
picture is valid.  This is best understood in the supersymmetric cases $k =
\pm 1$: the $T$-dual description has $n$ evenly spaced NS5-branes \OoguriWJ.
The sequences of transitions between non-supersymmetric
and supersymmetric four-dimensional ALE spaces
detailed in \S5\ (and presumably many others like
them) allow us to produce any supersymmetric
ALE space by closed-string tachyon condensation or
marginal deformation.  Using
the $T$-duality, we can restate this in terms
of NS5-branes.  Namely, any number of NS5-branes can be
obtained by condensation of modes in a non-supersymmetric
closed string background.

It is also interesting to look for a brane description
of the tachyonic starting point.  In particular, in the
case $\IC^2/\IZ_{2l(2l-1)}\to \IC^2/\IZ_{2(1)}$ one might
have expected that since the bosonic action is the same
as in a supersymmetric $\IC^2/\IZ_{2l}$
orbifold, the $T$-duality transformation
would produce a similar configuration of $2l$ NS5-branes.
However, the factor $(-1)^{\bf F}$ in the twist \rcaseone\
modifies the $T$-duality as described in \S6.1.  The $T$-dual circle is only
half as large, so there are only $l$ NS5-branes, while the bulk theory is
type 0 theory with a $(-1)^{\bf Q}$ twist around the $T$-dual circle.
The marginal deformations that we discussed descend from those of the
supersymmetric $\IZ_l$ theory, and so correspond to the positions of the $l$
NS5-branes.  The tachyons, in the sectors of odd $\IZ_2$ quantum symmetry,
are modes of the type 0 tachyon.

It would be interesting to pursue this type of dual description
of the non-supersym\-metric ALE orbifolds further.  It is straightforward to
apply the general $T$-duality transformation~\Buscher, but this results in
a smeared 5-brane solution and we do not know the localized form in general.

\subsec{Adding RR Flux}

A simple generalization is to add an RR Wilson line to the $\IC/\IZ_n$
orbifold,\foot{We thank
A. Strominger for discussions on this issue.}
$C_\theta = 1$ in coordinates where the identification is $\theta \sim
\theta+2\pi/n$. The net phase is then $2\pi/n$.  In M theory this corresponds
to an orbifold by a $2\pi/n$ rotation accompanied by a shift by $1/n$ around
the M theory  circle.  In a dual description where a linear
combination of the eleventh direction and the angular direction of
the orbifold is taken to be the M direction, this is a
fluxbrane~\DowkerBT\DowkerUP\CostaNW\GutperleMB\CostaIF\SaffinKY.  
Because of the factor of
$(-1)^{\bf F}$ in the orbifold, it is a fluxbrane in the type 0A
theory~\BergmanKM\ of strength $BR^2 = 1/n$, or a fluxbrane in the type IIA
theory of strength $BR^2 = 1 + 1/n$.

This duality is a strong-weak coupling duality, so that both
sides are not simultaneously weakly coupled.  However,
on the fluxbrane side the coupling varies with radial
distance from the origin, becoming weaker toward the origin.  If
we fix the string coupling to be $g_s < 1$ on the orbifold
side, on the fluxbrane side one has a region $r<l_sg_s^{1/3}n=nl_{P,11}$
near the origin
which has string coupling $g_s^{(f)}<1$.  For large $n$
and $g_s>0$,
this region can cover many string lengths.  We will
study the predictions of our results combined with
the conjectured orbifold/fluxbrane duality for
decay of the Type 0A tachyon in this region.
In our analysis in the bulk of this paper, we
worked in the classical string limit. As
we have just learned, in order to
dualize to a fluxbrane side with a significant region
of weak coupling near the origin, we must relax
this limit somewhat, and consider a nonvanishing
orbifold string coupling, though we can keep it weak.
For the remainder of this section, we will assume
that the decay process we studied proceeds similarly
at weak but nonzero coupling.

RR field strengths
couple to extra powers of $g_{\rm s}$ in the action and so
for weak string coupling they have only a
small effect on the tachyon decay process we have studied.  The decay
will proceed as we have described, with the RR flux ultimately dispersing when
we reach the flat space endpoint.  For the partial decay $n \to n-2$, the
outgoing pulse must contain a negative RR flux $2\pi({1\over n} -
{1 \over n-2})$. In the dual fluxbrane, the flux near the
origin {\it increases} in the transition, from
$1\over n$ to $1 \over n-2$.  According to the conjectured duality
dictionary in
\CostaNW\GutperleMB, this addition of flux takes the $0A$ theory
closer to the flat space IIA theory.

Therefore, by assuming the dualities described in 
\CostaNW\GutperleMB\CostaIF,
and combining them with our results on classical tachyon
decay in orbifolds, we predict that the type 0A tachyon in ten
dimensions
decays toward the flat ten-dimensional IIA vacuum.  This
agrees with the conjecture for the fate of the Type 0A
tachyon made in \GutperleMB\ based on extrapolating
to a regime where {\it non-perturbative} decays from
0A to IIA occur \WittenGJ\DowkerGB.
Our route to this conclusion is somewhat more direct, as we use
our classical results on tachyon decays in orbifolds rather
than non-perturbative instanton effects.  However, these statements
are still predicated on the conjectural non-supersymmetric
strong-weak coupling duality \BergmanKM\CostaNW\ assumed in \GutperleMB.
Therefore we regard this as a mild consistency check
of the proposal that the type 0A tachyon drives the theory
to the type IIA vacuum.

\newsec{Conclusions}

In this paper we have exhibited strong evidence that tachyonic
non-supersymmetric ALE spaces decay to supersymmetric ALE spaces
(including flat space).
There are several interesting lessons and directions for
future work that emerge from our analysis.

On the theoretical side, as we have emphasized at various
points, our results are rather similar to ones
that emerge in the study of open string tachyon condensation and
its relation to unstable brane annihilation.
It would be very interesting to understand how far
the analogy between twisted strings and open strings goes.
Is there a notion of confinement of twisted strings into
ordinary untwisted closed strings?  Is there a simplification
of closed string field theory if one focuses on twisted states
and regards untwisted strings as derivative degrees of
freedom obtained in internal
legs of the diagrams?  What does the similarity between
closed string and open string processes say about the
extent of applicability of K-theoretic techniques as a
function of $g_s$?

We should reemphasize, as discussed
in the introduction, that there is a similar puzzling
issue in the two cases. Namely as in the open string
case, our results point to the
need for a strictly classical stringy mechanism, different from the
Higgs mechanism, for lifting gauge
bosons living on decaying defects.
It is perhaps a clue that the disappearance of these
gauge bosons and the other phenomena we have observed
occurs in the closed string as well as open
string context:  whatever the physics is that gives rise
to these processes, it is not tied uniquely to the open string perturbation
expansion since it arises for twisted closed strings as well.

Another related lesson is the existence of a large class of
non-supersymmetric
configurations which, while unstable, do not ``decay to nothing'',
as a class of non-SUSY models without massless fermions
are known to do \WittenGJ\DowkerGB\FabingerJD.
Instead, they decay via a relatively well-controlled
weakly coupled process to stable supersymmetric configurations.
It would be very interesting to understand the fate of {\it compact}
non-supersymmetric orbifolds of the superstring with massless
fermions, particularly since as we discussed the time-dependent physics
in the compact case is very similar to that of a cosmology heading toward
a big crunch singularity.

In terms of model-building, these results, while mostly
negative for supersymmetry breaking, at least may help
direct attention to more stable possibilities than geometrical orbifolds
for breaking SUSY.  The fact that the noncompact models
decay to SUSY spaces provides a new indication of the
intrinsic role of SUSY within the theory.
Again, the question of the fate of
compact examples which are most relevant for phenomenological
model-building is still open.

Finally, it would be interesting to extend these results to other
cases, such as
intersecting ALE spaces probed by different combinations
of D-branes, and the type I theory.  In particular, dualities
suggest that the case of singular ALE spaces intersecting at
angles introduces novel phenomena \uranga, and it will
be interesting to see if our techniques in this paper
can provide insight into this case (or a deformation of it).

Noncompact tachyonic orbifolds of the heterotic string may
have a similar fate to those we discussed here, but in that case there are
no D-brane probes available to study the substring regime.
The heterotic case will
require an understanding of the dynamics of the vector bundle
formed by the gauge bosons as well as the configuration
of dilaton and metric.  Under RG flow the cases with a standard
embedding of the orbifold action into the gauge group will
behave as our models here; it would be interesting to study
also the time-dependent on-shell spacetime solutions in the heterotic
string.

\bigskip

\noindent{\bf Acknowledgements}

We would like to thank M. Ro\v cek for many useful and enjoyable discussions
on this topic.  We would also like to thank M. Berkooz, D. Gross,
G. Horowitz, S. Kachru, P. Kraus, E. Martinec, A. Strominger,
and W. Taylor for very
useful discussions.  This work was supported in part
by the NSF under grant numbers PHY97-22022 and PHY99-07949.  A. A.
and E.S. would like to thank the hospitality of the Institute for Theoretical
Physics at UCSB where this work was initiated. A.A. is also supported in part
by an NSF Graduate Fellowship, and   A.A. and E.S. by the DOE (contract
DE-AC03-76SF00515 and OJI) and the Alfred P. Sloan Foundation.

\listrefs

\end